\documentclass[12pt]{article}

\ifx\pdfoutput\undefined
\usepackage[dvips,bookmarks]{hyperref}
\else
\usepackage{hyperref}
\fi
\hypersetup{colorlinks=false,bookmarksopen,bookmarksnumbered,citecolor=blue,pdfstartview=FitH}

\usepackage{latexsym}
\usepackage{amssymb,amsfonts,amsmath}
\usepackage{graphicx}
\usepackage{indentfirst}
\usepackage{bbm}

\parskip=\medskipamount

\newcommand {\cA}{{\cal A}}
\newcommand {\cB}{{\cal B}}
\newcommand {\cC}{{\cal C}}
\newcommand {\cD}{{\cal D}}
\newcommand {\cE}{{\cal E}}

\newcommand {\cH}{{\cal H}}

\newcommand {\cL}{{\cal L}}

\newcommand {\cN}{{\cal N}}

\newcommand {\cP}{{\cal P}}

\def\a{\alpha}
\def\b{\beta}

\def\d{\delta}
\def\e{\epsilon}

\def\g{\gamma}

\def\k{\kappa}
\def\l{\lambda}
\def\m{\mu}
\def\n{\nu}

\def\q{\theta}

\def\s{\sigma}

\def\x{\xi}
\def\z{\zeta}
\def\D{\Delta}

\def\J{\Psi}
\def\L{\Lambda}
\def\O{\Omega}

\def\Q{\Theta}
\def\S{\Sigma}
\def\U{\Upsilon}
\def\X{\Xi}

\def\tr{{\rm tr}}

\def\ri{{\rm i}}
\def\re{{\rm e}}

\newcommand{\ve}{\varepsilon}                            

\newcommand{\ab}{{\a\b}}

\newcommand{\hf}{\frac12}

\newcommand{\vf}{\varphi}

\newcommand{\be}{\begin{equation}}
\newcommand{\ee}{\end{equation}}
\newcommand{\bea}{\begin{eqnarray}}
\newcommand{\eea}{\end{eqnarray}}
\newcommand{\non}{\nonumber}
\newcommand{\ba}{\begin{array}}
\newcommand{\ea}{\end{array}}

\newcommand{\au}{\underline{a}}

\newcommand{\bm}[1]{\mbox{\boldmath$#1$}}

\def\double #1{#1{\hbox{\kern-2pt $#1$}}}

\newcommand{\ha}{{\hat{a}}}
\newcommand{\hb}{{\hat{b}}}

\newcommand{\hal}{{\hat{\a}}}
\newcommand{\hbe}{{\hat{\b}}}
\newcommand{\hga}{{\hat{\g}}}
\newcommand{\hde}{{\hat{\d}}}

\newcommand{\sSp}{\mathsf{Sp}}
\newcommand{\sSU}{\mathsf{SU}}
\newcommand{\sSL}{\mathsf{SL}}
\newcommand{\sGL}{\mathsf{GL}}
\newcommand{\sSO}{\mathsf{SO}}
\newcommand{\sU}{\mathsf{U}}

\newcommand{\sUSp}{\mathsf{USp}}

\newcommand{\sOSp}{\mathsf{OSp}}

\newcommand{\sO}{\mathsf{O}}

\newcommand{\bsubeq}{\begin{subequations}}
\newcommand{\esubeq}{\end{subequations}}

\newcommand{\rd}{\mathrm d}
\begin{document}

\begin{titlepage}
\begin{flushright}
June, 2014
\end{flushright}
\vspace{5mm}

\begin{center}
{\Large \bf Superconformal structures on the three-sphere}
\\
\end{center}

\begin{center}

{\bf Sergei M. Kuzenko${}^{a}$ and D. Sorokin${}^{b}$}

\footnotesize{
${}^{a}${\it School of Physics M013, The University of Western Australia\\
35 Stirling Highway, Crawley W.A. 6009, Australia}}  ~\\
\vspace{5mm}

\footnotesize{
${}^{b}${\it INFN, Sezione di Padova, 35131 Padova, Italy}}\\[1mm]

\end{center}
\vspace{5mm}

\begin{abstract}
\baselineskip=14pt
With the motivation to develop superconformal field theory
on $S^3$, we introduce a $2n$-extended supersphere $S^{3|4n}$,
with $n=1,2,\dots$, as a homogeneous space of the three-dimensional
Euclidean superconformal group $\sOSp(2n|2,2)$ such
that its bosonic body is $S^3$.
Supertwistor and bi-supertwistor realizations of $S^{3|4n}$ are derived.
We study in detail the $n=1$ case,  which is unique in the sense that
the $R$-symmetry subgroup $\sSO^*(2n)$
of the superconformal group is compact only for $n=1$.
In particular, we show that the $\sOSp(2|2,2)$  transformations preserve
the chiral subspace of $S^{3|4}$.
Several supercoset realizations of $S^{3|4n}$ are presented.
Harmonic/projective extensions of the supersphere by auxiliary bosonic fibre directions are  sketched.
\end{abstract}
\vspace{0.5cm}

\vfill
\end{titlepage}

\newpage
\renewcommand{\thefootnote}{\arabic{footnote}}
\setcounter{footnote}{0}

 \tableofcontents


\numberwithin{equation}{section}



\section{Introduction}
Recently, there has been an interest (see, e.g., \cite{KWY,Jafferis,Hama:2010av})
in superconformal field theories on a three-dimensional ($3D$) sphere,
mostly motivated by the study of their quantum features with the use of localization techniques.
In addition to the issues raised in \cite{KWY,Jafferis,Hama:2010av} and related papers,
it is also of interest to study correlation functions in superconformal
field theories on $S^3$, and a superspace setting appears to be most suitable
to address this goal. An $\cN=2$ superspace formalism has been developed to describe
$\cN$-extended supersymmetric gauge theories on $S^3$ \cite{SS},
but superconformal aspects of these and more general theories have not been
studied in the Euclidean superspace framework
so far.\footnote{The construction of
$\cN=2$ supersymmetric theories on $S^3$ \cite{SS} is similar to that
of the off-shell (2,0) supersymmetric field theories in $AdS_3$
given in \cite{KT-M11}. In general, supersymmetric field theory in $AdS_3$
has so far been developed to a greater degree of completeness
than its Euclidean $S^3$ counterpart. The supersymmetric extensions of $AdS_3$
were constructed in \cite{KT-M11,KLT-M12} and are known as the
$(p,q)$ AdS superspaces, where
$p \geq q$ are non-negative integers.
For all types of $\cN =3$ and $\cN= 4$ AdS supersymmetry,
where $\cN=p+q$, general off-shell supersymmetric
field theories were constructed in a manifestly supersymmetric approach
and also reformulated in (2,0) AdS superspace \cite{KLT-M12,BKT-M,KT-M14}.}

This paper is designed to be one of  a series
devoted to off-shell superconformal field theories on $S^3$
and is aimed at setting a geometric stage for their further study.
We introduce a $2n$-extended supersphere $S^{3|4n}$,
with $n=1,2,\dots$, as a homogeneous space of the $3D$ Euclidean
superconformal group,
$\sOSp(2n|2,2)$, with the property
that the bosonic body of $S^{3|4n}$ is the three-sphere.\footnote{The supersphere $S^{3|4n}$ has
$4n$ Grassmann-odd directions that are parametrized by $2n$ two-component spinor coordinates.}
Supertwistor and bi-supertwistor realizations of $S^{3|4n}$ are derived.
To some extent, these realizations are analogous to those of 3D and 4D
compactified Minkowski superspaces $\overline{\mathbb M}{}^{3|2\cN}$
and $\overline{\mathbb M}{}^{4|4\cN}$, respectively, described in detail in
\cite{KPT-MvU,K06,K12}. However, the Euclidean case turns out to have new
nontrivial features.

This paper is organized as follows. In section 2, we collect the main definitions
concerning the $3D$ Euclidean conformal and superconformal groups.
In section 3, we describe the twistor and bitwistor realizations of $S^3$
as a warm-up for the subsequent supersymmetric constructions.
The supertwistor and bi-supertwistor realizations of the
$\cN=2n$ extended supersphere
$S^{3|4n}$ are presented in section 4.  The specific features of the
$\cN=2$ supersphere are analysed in section 5.
Several supercoset realizations of  $S^{3|4n}$ and
flat Euclidean superspace $\mathbb E^{3|4n}$
are given in section 6. The main body of the paper is accompanied by three
appendices. In appendix A, we review two different matrix realizations
for each of the groups  $ \sSp (2n, {\mathbb R} )$ and $\sSO^*(2n)$.
Appendix B is a brief review of the Veblen-Dirac
construction of pseudo-Euclidean conformal spaces $\overline{\mathbb E}^{s,t}$.
In appendix C, we sketch the construction of harmonic/projective extensions of
$S^{3|4n}$ by auxiliary bosonic variables.


\section{3D Euclidean  (super)conformal groups}

In this section we define the conformal and  superconformal groups
in three Euclidean dimensions.

\subsection{The conformal group}

The conformal group of both the three-sphere $S^3$ and
the Euclidean three-plane ${\mathbb E}^3$ is $\sSO (4,1)$.
The same group is also the isometry group of four-dimensional
($4D$) de Sitter space $dS_4$.
Its connected component  $\sSO_0 (4,1)$
 is locally isomorphic\footnote{The group  $\sUSp(2,2)$ is
 a two to one  covering group of $\sSO_0 (4,1)$.}
to the $dS_4$ spin group $\sUSp(2,2)$  defined by
\bea
\sUSp(2,2) = \sSU (2,2) \bigcap \sSp (4, {\mathbb C})_{ dS}~.
\eea
Here $\sSU(2,2)$ is a two to one  covering group of the connected component
$\sSO_0 (4,2)$ of the conformal group  of four-dimensional Minkowski space
${\mathbb M}^{4} ={\mathbb E}^{3,1}$,
\bea
\sSU(2,2) := \left\{ g \in \sSL (4,{\mathbb C}) ~,
\quad g^\dagger I g = I~,
\quad
I=\left(
\begin{array}{cc}
{\mathbbm 1}_2   & 0  \\
0 &     -{\mathbbm 1}_2
\end{array}
\right) \right\}~.
\eea
In the notation of appendix \ref{App0}, the matrix $I$ is $ I_{2,2} $.
The group $\sSp (4, {\mathbb C})_{dS}$ is simply the symplectic
group $\sSp (4, {\mathbb C})$ in the following realization:
\bea
\sSp (4, {\mathbb C})_{dS}  := \left\{ g \in \sGL (4,{\mathbb C}) ~,
\quad g^{\rm T} \L g = \L~,
\quad
\L =\left(
\begin{array}{cc}
\s_2   & 0  \\
0 &     -\s_2
\end{array}
\right) \right\}~,
\eea
where $\s_2$ is the second Pauli matrix. The matrix $\L$  satisfies the properties
\bea
\L^\dagger = -\L^{\rm T} = \L ~, \qquad
\L^2 = {\mathbbm 1}_4~.
\eea

It is instructive to compare the $dS_4$ spin group, $\sUSp(2,2)$,
with the one corresponding to  $4D$ anti-de Sitter space $AdS_4$,
$\sSp (4, {\mathbb R}) $.
 The latter is a two to one covering group of the
 connected component $\sSO_0(3,2)$ of the
isometry group of $AdS_4$.
As shown in  appendix \ref{App0}, this group can equivalently be realized
as a subgroup of $\sSU(2,2)$. This  follows from the isomorphism
\bea
\sSp (4, {\mathbb R}) \cong  \sSU (2,2) \bigcap \sSp (4, {\mathbb C})_{ AdS}~,
\eea
where $\sSp (4, {\mathbb C})_{ AdS}$ stands for  the symplectic
group $\sSp (4, {\mathbb C})$ in the following realization:
\bea
\sSp (4, {\mathbb C})_{ AdS}  := \left\{ g \in \sGL (4,{\mathbb C}) ~,
\quad g^{\rm T} J g = J~,
\quad
J =\left(
\begin{array}{cc}
0   & {\mathbbm 1}_2   \\
-{\mathbbm 1}_2  &     0
\end{array}
\right) \right\}~.
\eea
In the notation of appendix \ref{App0}, the matrix $J$ is $ J_{2,2} $.

As will be demonstrated in section \ref{Section:Three-sphere},
the above matrix realization of $\sUSp (2,2)$ is most suitable to describe
the global action of the superconformal group on the sphere $S^{3}$.
However, in order to describe the conformal transformations in flat
Euclidean space ${\mathbb E}^{3}$, a different matrix realization
of $\sUSp (2,2)$ is more convenient.
It is obtained from the original realization by applying
the following similarity transformation:
\bea
g ~ & \to & ~
{\bm g} =
\S \, g\, \S^{-1} ~, \quad g \in \sUSp(2,2)~,
\eea
where we have introduced the orthogonal $4\times 4$ matrix
\bea
\S= \frac{1}{  \sqrt{2} }
\left(
\begin{array}{cr}
 {\mathbbm  1}_2   & - {\mathbbm 1}_2\\
{\mathbbm  1}_2 &   ~ {\mathbbm 1}_2
\end{array}
\right)~, \qquad \S^{\rm T} \,\S= {\mathbbm 1}_4~.
\label{2.22}
\eea
In this realization, the elements of  $\sUSp (2,2)$ obey the constraints
\bea
{\bm g}^\dagger \bm I \bm g = \bm I~, \qquad
{\bm g}^{\rm T} \bm \L \bm g = \bm \L~,
\eea
where
\bea
\bm{I} = \S \, I\, \S^{-1}
= \left(
\begin{array}{cc}
 0 &  {\mathbbm 1}_2\\
{\mathbbm  1}_2 &    0
\end{array}
\right)~, \qquad
\bm \L = \S \, \L\, \S^{-1}
= \left(
\begin{array}{cc}
 0 &  \s_2\\
\s_2 &    0
\end{array}
\right)~.
\label{Metric3.27}
\eea

\subsection{The superconformal group}

 $\cN$-extended  superconformal group in three Euclidean dimensions is
\bea\label{osp}
\sOSp (2n|2,2) = \sSU(n,n |2,2) \bigcap \sOSp (2n| 4; {\mathbb C})~,
\qquad n=1,2, \dots~,
\eea
with $\cN=2n$.
It consists of $(2n|4) \times (2n|4)$  supermatrices
(with $A, D$ bosonic blocks and $B, C$ fermionic ones)
\bea
g = \left(
\begin{array}{c|c}
 A  & B\\
 \hline
C &    D
\end{array}
\right)~
\eea
constrained by
\begin{subequations} \label{slashX}
\bea
g^\dagger \X g &=& \X~, \qquad
\X = \left(
\begin{array}{c|c}
 \O   & 0 \\
 \hline
0 &     I
\end{array}
\right)~, \qquad \O^\dagger = - \O^{\rm T} = \O ~, \quad
\O^2 = {\mathbbm 1}_{2n}~, \label{slashX.a} \\
g^{\rm sT}
\U
g &=&
\U
~, \qquad
\U
= \left(
\begin{array}{c|c}
 {\mathbbm 1}_{2n}   & 0 \\
 \hline
0 &     \L
\end{array}
\right)~, \qquad
g^{\rm sT}=\left(
\begin{array}{c|c}
 A^{\rm T}  & C^{\rm T}\\
 \hline
-B^{\rm T}  &    D^{\rm T}
\end{array}
\right)~.
\eea
\end{subequations}
The bosonic subgroup of $\sOSp (2n|2,2) $ is $\sSO^*(2n) \times \sUSp (2,2)$.
Here  we define the group $\sSO^*(2n)$ by
\bea
\sSO^* (2n)  := \Big\{ \frak U \in \sGL (2n,{\mathbb C}) ~, \quad
{\frak U}^{\rm T} \frak U = {\mathbbm 1}_{2n} ~,
\quad {\frak U}^\dagger \O {\frak U}
= \O \Big\}
\label{B.5}~.
\eea
This definition of  $\sSO^*(2n)$ is equivalent to the standard one given
in appendix  \ref{App0}, eq. \eqref{AA.6}. Indeed,
it is always possible to choose $\O = \ri J_{n,n}$
by applying a similarity transformation.

The above supermatrix realization of $\sOSp (2n|2,2)$ is most suitable to consider
the global action of the superconformal group on the supersphere $S^{3|4n}$.
However, in order to describe superconformal transformations in flat
Euclidean superspace ${\mathbb E}^{3|4n}$, a different supermatrix realization
of $\sOSp (2n|2,2)$ is more useful. It is obtained from the above realization
by  applying a similarity transformation
\bea
g ~&\to &~ \bm g = \bm \S g {\bm \S}^{-1}~, \qquad
g \in \sOSp (2n|2,2)
\eea
associated with
the $(2n|4) \times (2n|4) $ supermatrix
\bea
\bm \S =
\left(
\begin{array}{c|c}
 {\mathbbm 1}_{2n}  & ~0 \\
 \hline
0 &     ~\S
\end{array}
\right)~,
\label{Sigma4.27}
\eea
where $\S$ is given  by \eqref{2.22}.
In the new realization, the group elements of $\sOSp (2n|2,2)$
obey the constraints
\bea
{\bm g}^\dagger \bm \X \bm g = \bm \X~, \qquad
{\bm g}^{\rm sT} \bm \U \bm g = \bm \U~,
\eea
where
\bea
\bm \X =  \bm \S\, \X \,{\bm \S}^{-1}
=\left(
\begin{array}{c|l c}
 \O   & ~0 & 0\\
 \hline
0 & ~   0 & {\mathbbm 1}_2 \\
0 &~ {\mathbbm 1}_2 & 0
\end{array} \right)~,
\qquad
\bm \U =  \bm \S \,\U \,{\bm \S}^{-1}= \left(
\begin{array}{c|l c}
 {\mathbbm 1}_{2n}   & ~0 & 0\\
 \hline
0 & ~   0 & \s_2 \\
0 &~ \s_2 & 0
\end{array} \right)~.
\label{3.30}
\eea

\subsection{The superconformal algebra}

Any element  $\cL$ of the superconformal algebra
$\frak{osp}(2n|2,2)$ obeys the equations
\begin{subequations}\label{salgebra}
\bea
\cL^\dagger { \X} +{ \X} \cL &=&0~, \\
\cL^{\rm  sT}  \U +\U \cL &=&0~,
\eea
\end{subequations}
which are the infinitesimal counterpart of \eqref{slashX}.
This gives a matrix realization of $\frak{osp}(2n|2,2)$.
Alternatively, the superconformal algebra may be defined
be specifying the corresponding (anti)commutation relations of its generators,
and without resorting to any particular matrix realization.

The superalgebra $\frak{osp}(2n|2,2)$ is formed by the generators
$L_{\hat a\hat b}=-L_{\hat a\hat b}$
of $\sSp(2,2)  \cong \sSO (4,1) /{\mathbb Z}_2$ ($\hat a,\hat b=0,1,2,3,4$),
the generators $T^{ij}=-T^{ji}$ of $\sSO^*(2n)$ $(i,j=1,\cdots 2n)$ and $8n$ supercharges $\mathcal Q_{\hat\alpha}^i$ $(\hat\alpha=1,2,3,4)$.\footnote{Note that $\frak so(4,1)$ generates the isometries of $dS_4$ space, and the superalgebra $\frak{osp}(2n|2,2)$ is a unique superextension of this algebra (see e.g. \cite{Lukierski:1981ht,Lukierski:1983qc}). $4D$ supergravity theories based on the $n=1$ $dS_4$ superalgebra were shown \cite{PvNS,LN} to contain ghosts.}
The defining anti-commutation relation of $\mathcal Q_{\hat\alpha}^i$ is (see e.g. \cite{Lukierski:1983qc})
\be\label{ospn22}
\{\mathcal Q_{\hat\alpha}^i,\mathcal Q^j_{\hat\beta}\}=\delta^{ij}\gamma^{\hat a\hat b}_{\hat\alpha\hat\beta}\,L_{\hat a\hat b}+ C_{\hat\alpha\hat\beta} T^{ij}\,.
\ee
Here
$C= (C_{\hat\alpha\hat\beta})$
is a charge conjugation matrix,
$C_{\hat\alpha\hat\beta}=-C_{\hat\beta\hat\alpha}=-C^{\hat\alpha\hat\beta}$,
which we choose  to be $C=\ri\Lambda$
and which is used to raise and lower the spinor indices
$$\mathcal Q^{i\hat\alpha}=C^{\hat\alpha\hat\beta}\mathcal Q^{i}_{\hat\beta}~, \qquad \mathcal Q^i_{\hat\alpha}= C_{\hat\alpha\hat\beta}\mathcal Q^{i\hat\beta}~.$$
The  $D=5$ gamma-matrices
$\gamma_{\hat a}=(\gamma_{\hat a}{}^{\hat\alpha}{}_{\hat\beta})$
obey the anti-commutation relation
\be\label{Galgebra}
\{\gamma_{\hat a},\gamma_{\hat b}\}
=-2\eta_{\hat a\hat b} {\mathbbm 1}_4~,
\ee
where
$\eta_{\hat a\hat b}={\rm diag}(-,+,+,+,+) $ is the Minkowski metric.
These matrices have the standard properties
\be
\gamma^{\dagger}_{ \hat a}= \gamma_0\gamma_{\hat a}\gamma_0~,
\qquad
(\gamma_{\hat a})^{\rm T}
=C\gamma_{\hat a}C^{-1}~,
\ee
which imply that
$\gamma^{\hat a}_{\hat\alpha\hat\beta}\equiv (C\gamma^{\hat a})_{\hat\alpha\hat\beta}=-(C\gamma^{\hat a})_{\hat\beta\hat\alpha}$ are antisymmetric and $\gamma^{\hat a\hat b}_{\hat\alpha\hat\beta}
\equiv (C\gamma^{\hat a\hat b})_{\hat\alpha\hat\beta}=(C\gamma^{\hat a\hat b})_{\hat\beta\hat\alpha}$ are symmetric matrices.
A convenient representation for
$\gamma_{\hat a}$
is
\bea\label{gamma}
\gamma_0 = \left(
\begin{array}{cc}
 {\mathbbm 1}_2  & 0\\
0 &    -{\mathbbm  1}_2
\end{array}
\right)= I~, \quad \gamma_a = \left(
\begin{array}{cc}
 0 &  \ri \s_a \\
\ri \s_a &    0
\end{array}
\right)~, \quad
\gamma_4 = \left(
\begin{array}{cc}
 0 &  {\mathbbm 1}_2\\
-{\mathbbm  1}_2 &    0
\end{array}
\right)~.
\eea

The superconformal generators $\mathcal Q^i$ are symplectic--Majorana spinors
\be\label{sM}
{\mathcal Q}^i=\ri\Omega^{ij}C(\bar {\mathcal Q}^j)^T\,,\qquad \bar {\mathcal Q}^i=({\mathcal Q}^i)^\dagger\gamma_0\,,
\ee
where $\Omega^{ij}=-\Omega^{ji}$ is an $\sSO^*(2n)$ invariant symplectic form
defined in \eqref{slashX.a}.
Note that the $\sSO^*(2n)$ indices are raised and lowered by the orthogonal unit metric $\delta^{ij}$.

Upon splitting the $D=5$ indices $\hat a,\hat b$ into Euclidean $D=3$ indices $a,b=1,2,3$ and the rest,  for instance $\hat a=(0,a,4)$,
we get the following $D=3$ adapted form of the $\sOSp(2n|2,2)$ algebra
\be\label{D3osp2n22}
\{{\mathcal Q}^i,{\mathcal Q}^j\}=\delta^{ij}(\gamma^{ab}\,L_{ab}+2\gamma^a\gamma^0\,L_{a0}+2\gamma^a\gamma^4\,L_{a4}+2\gamma^4\gamma^0\,L_{40})+ C T^{ij}\,,
\ee
where $L_{ab}$ generate an $\sSO(3)\cong \sSU(2)/{\mathbb Z}_2$ subgroup of the conformal group $\sSO(4,1)$ and $L_{a4}$ can be associated with the operators that generate translations in $S^3$, i.e. $L_{ab}$ and $L_{a4}$ form the $\sSO(4)  \cong
\big(\sSU(2)\times \sSU(2)\big)/{\mathbb Z}_2$ isometry of $S^3$, while $K_{a}=\ri(L_{a0}-L_{a4})$ and $L_{40}$ generate, respectively, the conformal boosts and dilatations of $S^3$.

The translations in a flat Euclidean $D=3$ space are generated by
$P_{a}=\ri (L_{a0}+L_{a4})$, while the flat space conformal boosts are generated by $K_{a}=\ri(L_{a0}-L_{a4})$. Note that $[P_a,P_b]=0=[K_a,K_b]$.


\section{The three-sphere as a conformal space}\label{Section:Three-sphere}

In this section we present twistor and bitwistor realizations for
the three-sphere.\footnote{The twistor and bitwistor realizations
for $4D$ conformal spaces
$\overline{\mathbb E}^{3,1} \equiv \overline{\mathbb M}^{4}$
and $\overline{\mathbb E}^{4,0} \equiv S^4$ (see appendix B for more details)
were given by Veblen in 1933 \cite{Veblen} who used
the Pl\"ucker-Klein correspondence.
He introduced the term ``spin-space''
for what nowadays is known as ``twistor space.'' Dirac learnt his realization
\cite{Dirac} of the conformal space $\overline{\mathbb M}^{4}$,
which is reviewed in appendix B, from Veblen
as acknowledged in \cite{Dirac}.}

\subsection{Twistor realization of the three-sphere}\label{subsection2.1}

Introduce two $\sUSp(2,2) $ invariant inner products on
  ${\mathbb C}^4$:
\begin{subequations} \label{2.6}
\bea
\langle S| T \rangle_{I}  &:=& S^\dagger \,{I } \, T
= \overline{S_{\hat \a} } I^{\hat \a \hat \b } T_{\hat \b}
~ , \label {2.6a} \\
\langle  S |T \rangle_{\L} &:=& S^{\rm T} \, \L \,T
= S_{\hat \a}  \L^{\hat \a \hat \b } T_{\hat \b}
~, \label{2.6b}
\eea
\end{subequations}
for any  $T,S \in {\mathbb C}^4$. We will refer to this space as twistor space,
and its elements will be called twistors.
A twistor is viewed as a column vector
\bea
T = (T_{\hal}) =\left(
\begin{array}{c}
 f_\a \\
  g_\b
  \end{array}
\right)~,\eea
with the two-component  spinors $f_\a$ and $g_\b$ being complex.

Consider the space of all two-planes  in ${\mathbb C}^4$ known as
the Grassmannian $G_{2,4}({\mathbb C})$.
Any two-plane is determined by its basis, i.e. by two linearly independent
twistors $T^\m$, with $\m=1,2$.
Such a  basis  $\{T^\m\}$ is defined only modulo
the equivalence relation
\be
\{ T^\m \} ~ \sim ~ \{ \tilde{T}^\m \} ~, \qquad
\tilde{T}^\m = T^\n\,R_\n{}^\m~,
\qquad R \in \sGL(2,{\mathbb C}) ~.
\label{nullplane2}
\ee
Equivalently,  the Grassmannian
$G_{2,4}({\mathbb C})$ can be thought of
as consisting of all
$4\times 2$  complex matrices of rank two,
\bea
( T^1 ~T^2 )=\left(
\begin{array}{c}
 F\\  G
\end{array}
\right) ~,
\label{two-plane}
\eea
where the $2\times 2$ matrices  $F$ and $G$ are
defined modulo the equivalence relation
\bea
\left(
\begin{array}{c}
 F\\  G
\end{array}
\right) ~ \sim ~
\left(
\begin{array}{c}
 F\, R\\  G\,R
\end{array}
\right) ~, \qquad R \in \sGL(2,{\mathbb C}) ~.
\label{ER1}
\eea

Let $\frak S$ denote the subspace of  $G_{2,4}({\mathbb C})$
consisting of all two-planes in ${\mathbb C}^4$ that are
null with respects to the two inners products \eqref{2.6}.
For any two-plane belonging to $\frak S$, it holds that
\bea
\langle T^\m | T^\n \rangle_{I} =0~, \qquad
\langle T^\m | T^\n  \rangle_{\L} =0~, \qquad \m, \n =1,2
\label{2.110}
\eea
or, equivalently,
\begin{subequations}
\bea
F^\dagger F - G^\dagger G &=& 0~, \label{2.11a} \\
F^{\rm T} \s_2 F - G^{\rm T} \s_2 G &=& 0~. \label{2.11b}
\eea
\end{subequations}

It is known that the space of all two-planes in ${\mathbb C}^4$
under the null condition \eqref{2.11a} is  compactified 4D Minkowski
space,
$\overline{\mathbb M}^4 =(S^{3} \times S^1)/{\mathbb Z}_2
$, see e.g. \cite{K06}.
As shown in \cite{K06}, the conditions that the $4 \times 2$ matrix \eqref{two-plane}
has rank two and obeys \eqref{2.11a} imply that
\bea
\det F \neq 0 \quad \mbox{and}  \quad \det G \neq 0~.
\eea
The equivalence relation \eqref{ER1} tells us that
\bea
\left(
\begin{array}{c}
 F\\  G
\end{array}
\right) ~ \sim ~\left(
\begin{array}{c}
 h\\  {\mathbbm 1}_2
\end{array}
\right) ~.
\label{2.13}
\eea
Now the conditions \eqref{2.11a} and \eqref{2.11b} imply, respectively,
\begin{subequations} \label{2.14}
\bea
h^\dagger h &=& {\mathbbm 1}_2 \quad \Longrightarrow \quad
h \in \sU(2) ~;\\
h^{\rm T} \s_2 h & =& \s_2 \quad \Longrightarrow \quad
\det h = 1~.
\eea
\end{subequations}
We conclude that ${\frak S}$ may be identified with the group manifold
$ \sSU(2) = S^3$.

Given a group element
\bea
g=\big(g_{\hat \a}{}^{\hat \b}\big)= \left(
\begin{array}{cc}
 A  & B\\
C &    D
\end{array}
\right) \in \sUSp(2,2) ~,
\label{GE}
\eea
with  $A,B,C$ and $D$ some  $2\times 2$ matrices,
its action on $S^3$ is a fractional linear transformation
\bea
h ~\to ~ h' = (Ah +B )(Ch +D)^{-1}~.
\label{2.188}
\eea

\subsection{Real structure} \label{subsection2.2}

Twistors transform in the defining representation of $\sUSp(2,2)$.
Given a group element of $\sUSp(2,2)$, eq. \eqref{GE},
it acts on twistor space as
\bea
T_{\hat \a} ~\to ~ T'_{\hat \a}= g_{\hat \a}{}^{\hat \b} T_{\hat\b}~.
\eea
Let us also consider the dual of twistor space. Its elements
are complex row vectors $V = (V^{\hat \a}) = (v^\a, w^\b)$
possessing the $\sUSp(2,2)$ transformation law
\bea
V^{\hat \a} ~\to ~ V'^{\hat \a}= V^{\hat \b} (g^{-1} )_{\hat \b}{}^{\hat \a}~.
\eea
Since both inners products \eqref{2.6} are  $\sUSp(2,2)$ invariant, we conclude that
$S^\dagger I$ and $S^{\rm T}\L$ are dual twistors for any twistor $S$.
The dual of $T_\hal$ is defined to be
\bea
\bar T^\hal := \overline{T_\hbe} I^{\hbe \hal} = I^{\hal \hbe } \overline{T_\hbe} ~.
\eea
We also point out that  $\L^{\hat \a \hat \b } $ is an invariant tensor of
$\sUSp(2,2)$, and so is its inverse $\L^{-1} = (\L_{\hat \a \hat \b})$.
As a result, we can define a one-to-one anti-linear map  of twistor space
onto itself,
\bea
\star: ~T_{\hat \a} ~\to ~ \L_{\hat \a \hat \b} \,I^{\hat \b \hat \g}\, \overline{T_{\hat \g}}~,
\label{2.18}
\eea
for any twistor $T$. This map induces a well defined transformation on
the Grassmannian $G_{2,4}({\mathbb C})$,
\bea
\cP = \left(
\begin{array}{c}
 F\\  G
\end{array}
\right) ~\to ~ \star \cP= \L^{-1} I \left(
\begin{array}{c}
\bar  F\\ \bar  G
\end{array}
\right)~.
\label{2.19}
\eea
This transformation is well defined in the sense that
any two equivalent $4\times 2$ matrices
$\cP $ and $\cP R$, with $R \in \sGL(2,{\mathbb C})$,
are mapped into equivalent ones, $\star \cP$ and
$(\star \cP) \bar{R}$,
where $\bar R$ denotes the complex conjugate of $R$.

The map \eqref{2.18} is characterized by the property
$ \star \star = - {\mathbbm 1}_4$, and therefore it
cannot be used to define a complex conjugation on
twistor space.\footnote{The map  \eqref{2.18} does not allow us
to define real lines in the space ${\mathbb C}P^3$ of lines in twistor space.}
However, the map \eqref{2.19} may be seen to define an involution on
the space of all two planes  in twistor space, $\star \star = {\rm id}$.
Now consider any null two-plane defined by the relations \eqref{2.13}
and \eqref{2.14}. It is straightforward  to show that this two-plane is real
with respect to the involution introduced.

\subsection{Bitwistor realization}\label{Subsection2.3}

Let $T_\hal{}^\m $ be two linearly independent twistors that form a
basis of a two-plane in ${\mathbb C}^4$. We can associate with them a bitwistor
\bea
X_{\hal \hbe} := T_{\hal}{}^\m T_{\hbe}{}^\n \ve_{\m \n} = - X_{\hbe \hal}
~, \qquad \ve_{\m \n }= -\ve_{\n \m}~, \qquad
\ve_{12} =-1~
\label{2.24}
\eea
and its dual
\bea
\bar X^{\hal \hbe} :=
\ve_{\m\n} \bar T^{\m \hal } \bar T^{\n \hbe}
=I^{\hal \hga} I^{\hbe \hde} \overline{X_{\hga \hde}}~,
\qquad  \bar T^{\m \hal } :=  \overline{ T^{ \hal \m}}~,
\eea
with $I= (I^{\hal \hbe})$.
In terms of $X_{\hal \hbe}$,
the equivalence relation  \eqref{nullplane2} turns into
\bea
X_{\hal \hbe}~\sim ~c X_{\hal \hbe}~, \qquad c \in {\mathbb C} \setminus \{ 0\}~.
\label{2.38}
\eea
In the case that the twistors $T_\hal{}^\m $ describe a null two-plane,
eq. \eqref{2.110}, the corresponding bitwistor $X_{\hal \hbe}$ has the
following algebraic properties:
\begin{subequations} \label{2.39}
\bea
X_{[\hal \hbe} X_{\hga  \hde ]} &=&0~,  \label{2.39a}
\\
\L^{\hbe \hal} X_{\hal \hbe}  &=&0~,    \label{2.39b} \\
\bar X^{\hal \hga} X_{\hga \hbe} &=&0~.  \label{2.39c}
\eea
\end{subequations}
As shown in subsection \ref{subsection2.2}, all null two-planes are real
with respect to the anti-linear map \eqref{2.18}. Recast in terms of $X_{\hal \hbe}$,
this property means the following:
\bea
\L_{\hal \hga} \L_{\hbe \hde} \bar X^{\hga \hde} ~\sim ~ X_{\hal \hbe}~.
\label{2.40}
\eea

The above discussion naturally leads us to an alternative  realization
of $S^3$ as the space of non-zero bitwistors
$X_{\hal \hbe}$ subject to the constraints \eqref{2.39} and defined
modulo the equivalence relation \eqref{2.38}.
Equivalence of this  {\it bitwistor realization} of $S^3$ to the twistor one
given in subsection \ref{subsection2.1}
can be proved in complete analogy to the case of
compactified 3D Minkowski space \cite{K12}.
Constraint \eqref{2.39a} means that $X_{\hal \hbe}$ is decomposable,
eq. \eqref{2.24}. The constraints  \eqref{2.39b} and  \eqref{2.39c}
prove to imply the null conditions \eqref{2.110}.

The bitwistor realization is intimately related to the Veblen-Dirac
realization of $S^3$, see appendix \ref{AppendixA}.
To see this, using the gamma--matrices \eqref{gamma}, we introduce a null five-vector
\bea
X_\ha :=   \gamma_\ha{}^{\hal\hbe} X_{\hal \hbe}~, \qquad \eta_{\ha \hb} X^\ha X^\hb =0~.
\eea
This vector is defined up to re--scalings  and, due to \eqref{2.40}, may be chosen to be real.
As a result, we arrive at the realization of $S^3$ described in appendix \ref{AppendixA}.


\subsection{Atlas on the  three-sphere} \label{Subsection2.4}

Let us switch to a new parametrization of the group $\sUSp(2,2)$ that is more convenient for describing the conformal transformations in ${\mathbb E}^3$.
This parametrization is obtained by applying
the similarity transformation
\begin{subequations}
\bea
g ~ & \to & ~
{\bm g} =
\S \, g\, \S^{-1} ~, \quad g \in \sUSp(2,2)~, \\
T ~ & \to & ~
{\bm T} =
\S \, T~, \quad T \in {\mathbb C}^4~,
\eea
\end{subequations}
with the matrix $\S$ given by \eqref{2.22}.
The matrices $I$ and $\L$,  which determine the inner products \eqref{2.6},
turn into those given by \eqref{Metric3.27},
while the two-plane turns into
\bea
\cP = \left(
\begin{array}{c}
 h\\  {\mathbbm 1}_2
\end{array}
\right) ~\to ~ \bm{\cP} = \frac{1}{\sqrt{2}} \left(
\begin{array}{c}
 h - {\mathbbm 1}_2\\ h+ {\mathbbm 1}_2
\end{array}
\right) ~.
\eea
The equations  $\det (h + \mathbbm{1}_2)= 0$
and $\det (h - \mathbbm{1}_2)= 0$,
with $h \in \sSU(2) $,
have unique solutions $h = - {\mathbbm 1}_2$
and $h =  {\mathbbm 1}_2$, respectively.
As a result, the sphere $S^3$ can be covered by two open charts,
$S^3 = U_{\rm N} \bigcup U_{\rm S} $. The north chart $U_{\rm N} $ is defined to
consist of all null two-planes  for which $\det (h + \mathbbm{1}_2)\neq 0$.
In this chart
\bea
\left(
\begin{array}{c}
 h - {\mathbbm 1}_2\\ h+ {\mathbbm 1}_2
\end{array}
\right) ~\sim ~
\left(
\begin{array}{c}
 \ri \,x_{\rm N} \\  {\mathbbm 1}_2
\end{array}
\right) ~, \qquad
\ri \,x_{\rm N} = \frac{h - {\mathbbm 1}_2}{h + {\mathbbm 1}_2}~.
\eea
Similarly, the south chart $U_{\rm S} $
is spanned by all null two-planes  with $\det (h - \mathbbm{1}_2)\neq 0$.
In this chart
\bea
\left(
\begin{array}{c}
 h - {\mathbbm 1}_2\\ h+ {\mathbbm 1}_2
\end{array}
\right) ~\sim ~
\left(
\begin{array}{c}
{\mathbbm 1}_2\\    \ri \,x_{\rm S}
\end{array}
\right) ~, \qquad
\ri \,x_{\rm S} = \frac{h + {\mathbbm 1}_2}{h - {\mathbbm 1}_2}~.
\eea
In the overlap of the two charts, $U_{\rm N} \bigcap U_{\rm S}$,
we have the transition function
\bea
x_{\rm S} = - x_{\rm N}{}^{-1} ~.
\eea

In the remainder of this subsection,
we  work in the north chart and denote the $2\times 2$
matrix $ x_{\rm N}$ simply by $x$. The null conditions
\eqref{2.14} imply that the matrix $x$ is constrained by
\bea
x^\dagger = x~,
\qquad x^{\rm T} = - \s_2 x \s_2 \quad \Longrightarrow \quad
x = \vec{x} \cdot \vec{\s} ~, \qquad \vec{x} \in {\mathbb R}^3~.
\eea
Thus we may think of $S^3$ as $ {\mathbb R}^3 \bigcup \{ \infty_{\rm N} \} $,
where ${\mathbb R}^3$ is identified with $U_{\rm N}$
and the point $\infty_{\rm N}$ is identified with the null two-plane
\bea
\bm{\cP}_{\infty_{\rm N}} = \left(
\begin{array}{c}
{\mathbbm 1}_2 \\  0
\end{array}
\right)   ~,
\eea
which corresponds to the origin of the coordinate chart $U_{\rm S}$.

In the new parametrization introduced, the conformal group
$\sUSp(2,2)$ consists of all $4\times 4$ matrices $g$ of the form:
\bea
\bm g= \left(
\begin{array}{cc}
 \cA  & \cB\\
\cC &    \cD
\end{array}
\right) ~, \qquad  {\bm g}^\dagger \bm{I} \bm g = \bm{I}~,
\qquad
\bm g^{\rm T} \bm \L \bm g = \bm \L~.
\eea
Given such a group element, $\bm g \in \sUSp(2,2)$,
it generates the following transformation on $S^3$:
\bea
\ri x ~\to ~\ri x' = ( \ri \cA x + \cB) (\ri \cC x + \cD)^{-1} ~.
\label{2.41}
\eea
The isotropy group of the point $\infty_{\rm N}$ consists of all matrices
of the form:
\bea
  \left(
\begin{array}{cc}
 {\mathbbm 1}_2  ~& \ri  {\frak b} \\
0 ~&    {\mathbbm 1}_2
\end{array}
\right)
 \left(
\begin{array}{cc}
 \re^{\hf \l} {\mathbbm 1}_2 & 0\\
0  &      \re^{-\hf \l}{\mathbbm 1}_2
\end{array}
\right)
 \left(
\begin{array}{cc}
    \frak R
    ~& 0
  \\
0 ~&
\frak R
\end{array}
\right) ~,
\qquad \l \in {\mathbb R}, \quad \frak R \in \sSU(2)~,
\label{2.43}
\eea
where we have denoted
${\frak b}:= \vec{b}\cdot\vec{\s}$, $\vec{b} \in {\mathbb R}^3$.
The parameters $\vec{b}$, $\l$ and $\frak R$ describe, respectively,
a translation, a dilatation and a rotation of Euclidean three-plane ${\mathbb E}^3$.
Transformations \eqref{2.43} with $\l=0$ span the connected isometry group
 of ${\mathbb E}^3$, $\mathsf{ISO }_0 (3)$.

The origin of $U_{\rm N}$,  $x=0$,
is the infinitely separated point $\infty_{\rm S}$
for  $U_{\rm S}$.
The isotropy group of this point consists of all matrices of the form:
\bea
 \left(
\begin{array}{cc}
 {\mathbbm 1}_2  ~&0 \\
 \ri  {\frak c} ~&    {\mathbbm 1}_2
\end{array}
\right)
 \left(
\begin{array}{cc}
 \re^{\hf \l} {\mathbbm 1}_2 & 0\\
0  &      \re^{-\hf \l} {\mathbbm 1}_2
\end{array}
\right)
 \left(
\begin{array}{cc}
\frak R   & ~0\\
 0   &    ~ \frak R
\end{array}
\right) ~, \qquad \l \in {\mathbb R}, \quad \frak R \in \sSU(2)~, \quad
\eea
with ${\frak c}:= \vec{c}\cdot\vec{\s}$, $\vec{c} \in {\mathbb R}^3$.
As follows from \eqref{2.41}, the parameter $\vec{c}$
generates a special conformal transformation of ${\mathbb E}^3$.


\section{The supersphere as a conformal superspace}

In this section we introduce a $2n$-extended supersphere $S^{3|4n}$
as a homogeneous space for the superconformal group $\sOSp (2n|2,2)$.
For this we develop supertwistor and bi-supertwistor
realizations for the supersphere.\footnote{The concept of
supertwistors was introduced by Ferber \cite{Ferber} within the framework  of
$4D$ conformal supersymmetry. The supertwistor realization for
compactified $4D$ $\cN$-extended Minkowski superspace
$\overline{\mathbb M}^{4|4\cN}$ was developed by Manin \cite{Manin}
and also Kotrla and  Niederle \cite{Niederle}. The bi-supertwistor
realization for the same superspace was first considered by Siegel
\cite{Siegel93,Siegel95}, although it naturally follows from Manin's
construction \cite{Manin}. See \cite{K06, K12} for modern descriptions
of these realizlations.}

\subsection{Supertwistors}

The supergroup $\sOSp (2n|2,2)$ naturally acts on the space of even
supertwistors
and also on the space of odd supertwistors. An arbitrary supertwistor
looks like
\bea
{ T} = ({ T}_A) = \left(
\begin{array}{c}
{ T}_i \\
{ T}_{\hal}
\end{array}
\right)
~, \qquad i = 1, \dots, 2n~.
\eea
In the case of even supertwistors, $ T_i$ is fermionic and $ T_\hal $ is bosonic.
In the case of odd supertwistors, $ T_i$ is bosonic and $ T_\hal $ is fermionic.
We introduce the parity function $\ve ( T )$ defined as:
$\ve ( T ) = 0$ if $ T$ is even, and $\ve ( T ) =1$ if $T $ is odd. We
also define
\bea
 \ve_A = \left\{
\begin{array}{c}
 1 \qquad A=i \\
 0 \qquad A=\hal
\end{array}
\right.{}~.
\non
\eea
Then the above definition can be rewritten as
\bea
\ve ( T_A) = \ve ( T ) + \ve_A \quad (\mbox{mod 2})~.
\eea
Even and odd supertwistors are called  pure.\footnote{This terminology
is natural within the framework of supervector spaces \cite{DeWitt,BK}
and should not be confused with Cartan's pure spinors \cite{Cartan}.}
The space of even supertwistors may be identified with ${\mathbb C}^{4|2n}$.

Supertwistors transform  in the defining representation
of  $\sOSp (2n|2,2)$,
\bea
T ~\to ~T' = g T~, \qquad g \in \sOSp (2n|2,2)~.
\eea
This transformation law implies that  the supergroup $\sOSp (2n|2,2)$ defined by \eqref{osp}--\eqref{slashX}
leaves invariant two inner products
 \begin{subequations} \label{3.7}
\bea
\langle  S |  T \rangle_\X &:=&  S^\dagger \,\X \,  T
= \overline{ S_A} \,\X^{AB} \, T_B
~ ,   \label{3.7a} \\
\langle   S |  T \rangle_{\U} &:=&
(-1)^{\ve_A +  \ve( S) \cdot \ve_A }S_A  \U^{AB} T_B~,
\label{3.7b}
\eea
\end{subequations}
for arbitrary pure supertwistors $ S$ and $ T$.
These inner products have the following fundamental properties:
 \begin{subequations}
\bea
\overline{ \langle  T_1 |  T_2 \rangle_\X } &=&
 \langle  T_2 |  T_1 \rangle_\X
~ ;  \\
\langle   T_1 |  T_2 \rangle_{\U} &=&
-(-1)^{\ve_1 \ve_2} \langle   T_2 |  T_1 \rangle_{\U}
~,
\eea
\end{subequations}
for arbitrary pure supertwistors $ T_1$ and $ T_2$.

A dual supertwistor
\bea
{ Z} = ({Z}^A) = \left(
{ Z}^i ,
{Z}^{\hal}
\right)
~, \qquad i = 1, \dots, 2n
\eea
transforms under  $\sOSp (2n|2,2)$ such that $Z^A T_A$ is invariant
for any supertwistor $T$,
\bea
Z~\to ~Z' = Zg^{-1}~,  \qquad g \in \sOSp (2n|2,2)~.
\eea
A dual supertwistor $Z$ is even (odd) if $Z^A T_A$ is a $c$-number
for any even (odd) supertwistor $T$.

Invariance of the inner product \eqref{3.7b} under $\sOSp (2n|2,2)$
tells us that
\bea
Z^A :=- (1)^{\ve_B +\ve(S) \ve_B} S_B \U^{BA}
= (-1)^{\ve(S) \ve_A} \U^{AB} S_B
\eea
is a  pure dual supertwistor. Conversely, given a  pure dual supertwistor $Z^A$,
the following object
\bea
S_A := (-1)^{\ve(Z) \ve_B} (\U^{-1})_{AB} Z^B
\label{3.11}
\eea
is a pure supertwistor. We emphasize that $\U^{AB}$ is an invariant tensor
of the superconformal group,
\bea
(g^{\rm sT})^A{}_C \U^{CD} g_D{}^B = \U^{AB}~, \qquad
(g^{\rm sT} )^A{}_B = (-1)^{ \ve_A \ve_B +\ve_B} g_B{}^A~,
\eea
for any group element $ g \in \sOSp (2n|2,2)$.

Since the inner product \eqref{3.7a} is invariant under
$\sOSp (2n|2,2) \subset $,
we observe that
\bea
\bar S ^A:=\overline{ S_B} \,\X^{BA}
\eea
is a dual supertwistor, for any pure supertwistor $S_A$.\footnote{ Eq.  \eqref{3.7a}
can be rewritten in the form $\langle  S |  T \rangle_\X = \bar S^A T_A$.}
In conjunction with our previous result \eqref{3.11},
this implies  the existence of a one-to-one map of supertwistor space
onto itself defined by
\bea
\star : ~ S_A ~\to~ (\star S)_A := (-1)^{ \ve_C +\ve(S)  \ve_C }
(\U^{-1})_{AB} \X^{BC} \overline{S_C}~,
\label{3.13}
\eea
for any pure supertwistor $S_A$.
This map is characterized by the property
\bea
\star \star = - {\mathbbm 1}_{2n|4}~,
\label{3.14}
\eea
which follows from
the observations that
the matrices $\O$ and $\L I$
(i) are purely imaginary;  and (ii) fulfill
the identities $\O^2 = {\mathbbm 1}_{2n}$
 and $(\L I)^2 ={\mathbbm 1}_4$.


\subsection{The supersphere} \label{Subsection3.2}

We define a $2n$-extended supersphere $S^{3|4n}$ to be the space
of all null and real two-planes in the space of even supertwistors
 ${\mathbb C}^{4|2n}$.
 In general, any two-plane in  ${\mathbb C}^{4|2n}$ is generated by two
 supertwistors $T^\m$ such that their bodies are linearly independent.
Equivalently, it may be described by a rank-two
$(2n|4)\times 2$ supermatrix
\bea
( T^\m )=\left(
\begin{array}{c}
 \Q\\
\hline
 F \\
 G
\end{array}
\right) ~, \qquad \m =1,2~,
\eea
which is defined modulo the equivalence relation
\bea
\left(
\begin{array}{c}
 \Q  \\  \hline
 F \\ G
\end{array}
\right) ~ \sim ~
\left(
\begin{array}{c}
 \Q\, R\\  \hline
  F\,R \\  G\,R
\end{array}
\right) ~, \qquad R \in \sGL(2,{\mathbb C}) ~.
\label{3.16}
\eea
Here $\Q$ is  a $2n \times 2$
fermionic matrix,
and
$F$ and $G$ are $2\times 2$
bosonic matrices.
The two-planes belonging  to  $S^{3|4n}$ are required to be
(i) null with respect to the two inner products \eqref{3.7};
and (ii) real with respect to the star-map \eqref{3.13}
modulo the equivalence relation \eqref{3.16}.
The null conditions are
\begin{subequations}\label{3.17}
\bea
\Q^\dagger \O \Q + F^\dagger F - G^\dagger G &=& 0 ~; \\
-\Q^{\rm T} \Q +F^{\rm T} \s_2 F - G^{\rm T} \s_2 G &=&0~.
\eea
\end{subequations}
As in the bosonic case, the first null condition implies
that $\det F \neq 0$ and $\det G \neq 0$.
As a result, the null two-plane can equivalently be described by a supermatrix
\bea
\cP =\left(
\begin{array}{c}
 \Q  \\  \hline
 \bm h \\
 {\mathbbm 1}_2
\end{array}
\right)
= \left(
\begin{array}{c}
 \Q_i{}^\b  \\  \hline
  {\bm h}_\a{}^\b  \\
 \d_\g{}^\b
\end{array}
\right)
~, \label{3.18}
\eea
where the null conditions \eqref{3.17} now read
\begin{subequations}\label{3.19}
\bea
\Q^\dagger \O \Q + {\bm h}^\dagger \bm  h &=& {\mathbbm 1}_{2} ~, \\
-\Q^{\rm T} \Q +{\bm h} ^{\rm T} \s_2 \bm h   &=&\s_2~.
\eea
\end{subequations}
The condition that  the two-plane \eqref{3.18}  is real under  \eqref{3.13}
amounts to
\begin{subequations}\label{3.20}
\bea
\overline{\Q}     &=& - \O \Q\s_2  ~,  \label{3.20a} \\
\overline{ \bm h}   &=&\s_2 \bm h \s_2~. \label{3.20b}
\eea
\end{subequations}
Eq. \eqref{3.20a} is a pseudo-Majorana condition.


\subsection{Bi-supertwistor realization}

The bitwistor realization of the three-sphere given in subsection
\ref{Subsection2.3} can naturally be generalized to the case of the supersphere.

Let $T_A{}^\m $ be two linearly independent even supertwistors
belonging to a two-plane in ${\mathbb C}^{4|2n}$.
We can associate with them a bi-supertwistor
\bea
X_{A B} := T_{A}{}^\m T_{B}{}^\n \ve_{\m \n} = - (-1)^{\ve_A \ve_B}X_{B A}~.
\eea
The equivalence relation \eqref{3.16} turns into
\bea
X_{AB}~\sim~ c X_{AB}~, \qquad c \in {\mathbb C} \setminus \{ 0\}~.
\label{bi-s24}
\eea
Using the dual supertwistors $\bar T^{\m \,A }: = \overline{ T_B{}^\m } \X^{BA}$
we define a dual bi-supertwistor as
\bea
\bar X^{AB} := \ve_{\m\n} \bar T^{\m\, A }\bar T^{\n \, B} = - (-1)^{\ve_A \ve_B}X^{B A}~.
\eea
The supermatrices $X=(X_{AB})$ and $\bar X=(\bar X^{AB})$ are related to each other as
\bea
\bar X^{AB} = -(-1)^{\ve_C} \X^{AC} (X^\dagger )_{CD} \X^{DB} ~,
\qquad (X^\dagger )_{AB} := \overline{X_{BA} }~.
\eea
In the case that $T_A{}^\m $ generate a null two-plane,
the associated bi--supertwistor $X_{AB}$ has the following properties:
\begin{subequations}\label{bi-s27}
\bea
X_{[AB} X_{CD\}} &=&0~,\\
\U^{BA} X_{AB} &=&0 ~,\\
\bar X^{AB} X_{BC}&=&0~.
\eea
\end{subequations}
In terms of $X_{AB}$, the reality conditions \eqref{3.20} take the form:
\bea
(\U^{-1})_{AC} (\U^{-1})_{BD} \bar X^{CD} \propto X_{AB}~.
\label{bi-s28}
\eea

The above consideration naturally leads to a new realization of
the supersphere $S^{3|4n}$. In the space of graded antisymmetric
supermatrices $X_{A B} = - (-1)^{\ve_A \ve_B}X_{B A}$,
we consider a surface $\frak L$
spanned by those supermatrices which (i) obey the algebraic constraints
\eqref{bi-s27}; (ii) satisfy the reality condition \eqref{bi-s28};
and (iii) have the property that
the  {\it body} of the bosonic block
${ X}_{\hal \hbe}$ defined by
\bea
X_{AB} =
\left(
\begin{array}{c | c}
 { X}_{ij} & { X}_{i \hbe }  \\
 \hline
{ X}_{\hal j}  & { X}_{\hal \hbe}
\end{array}
\right)
\eea
is a non-zero antisymmetric $4\times 4$ matrix.
It may be shown\footnote{The proof is analogous to the one given in \cite{K12}
in the Lorentzian case.}
that the quotient space of $\frak L$ with respect to
\eqref{bi-s24} is equivalent to $S^{3|4n}$.


\subsection{Atlas on the supersphere}

Now we introduce an atlas on $S^{3|4n}$ as a natural generalization
of the bosonic construction described in subsection \ref{Subsection2.4}.
A bi-product of our consideration in this subsection  will be a  formalism
to describe the superconformal transformations
in flat Euclidean superspace ${\mathbb E}^{3|4n}$.

 It is advantageous to introduce a new parametrization of
the superconformal group $\sOSp (2n|2,2)$
obtained by applying a
similarity transformation associated with
the $(2n|4) \times (2n|4) $ supermatrix \eqref{Sigma4.27}.
The similarity transformation is defined as
\begin{subequations}\label{MR4.27}
\bea
g ~&\to &~ \bm g = \bm \S g {\bm \S}^{-1}~, \qquad
g \in \sOSp (2n|2,2)~; \\
T ~&\to &~ \bm T = \bm \S T ~,
\eea
\end{subequations}
for any pure supertwistor $T$.

The null two-plane \eqref{3.18} turns into
\bea
\cP =\left(
\begin{array}{c}
 \Q  \\  \hline
 \bm h \\
 {\mathbbm 1}_2
\end{array}
\right) ~\to ~{\bm \cP} = \frac{1}{\sqrt{2}}
\left(
\begin{array}{c}
\sqrt{2}  \Q  \\
\hline
 \bm h -{\mathbbm 1}_2\\
 \bm h + {\mathbbm 1}_2
\end{array}
\right)~.
\eea
A natural  atlas on $S^{3|4n}$ consists of two charts,
$S^{3|4n} = U_{\rm N} \bigcup U_{\rm S} $,
where the open sets $U_{\rm N}$ and $U_{\rm S}$
are  defined by the conditions $\det(\bm h +{\mathbbm 1}_2) \neq 0$
and $\det(\bm h - {\mathbbm 1}_2) \neq 0$, respectively.
In the north chart,
the above two-plane is equivalently described by
\bea
\bm \cP ~\sim ~
\left(
\begin{array}{c}
 \bm \q_{\rm N} \\  \hline
  \ri {\bm x}_{\rm N} \\
 {\mathbbm 1}_2
\end{array}
\right) ~, \qquad \ri {\bm x}_{\rm N} :=
\frac{\bm h -{\mathbbm 1}_2} {\bm h + {\mathbbm 1}_2} ~, \qquad
\bm \q_{\rm N} :=\sqrt{2} \Q (\bm h +{\mathbbm 1}_2 )^{-1}~.
\label{3.34north}
\eea
In the south chart, the same two-plane is parametrized by
\bea
\bm \cP ~\sim ~
\left(
\begin{array}{c}
\bm \q_{\rm S} \\  \hline
  {\mathbbm 1}_2 \\
 \ri {\bm x}_{\rm S}
\end{array}
\right) ~, \qquad \ri {\bm x}_{\rm S} :=
\frac{\bm h +{\mathbbm 1}_2} {\bm h - {\mathbbm 1}_2} ~, \qquad
\bm \q_{\rm S} :=\sqrt{2} \Q (\bm h -{\mathbbm 1}_2 )^{-1}~.
\eea
In the overlap of the two charts, $U_{\rm N} \bigcap U_{\rm S} $,
we obtain the transition functions
\bea
 {\bm x}_{\rm S} = -  {\bm x}_{\rm N}{}^{-1}~, \qquad
\bm \q_{\rm S} = - \ri \,\bm \q_{\rm N}  \, {\bm x}_{\rm N} {}^{-1}~.
\eea

The point $\infty_{\rm S} \in U_{\rm N}$
labeled  by ${\bm x}_{\rm N}=0$ and ${\bm \q}_{\rm N} =0$
is infinitely separated from the point of view of  $U_{\rm S}$.
Similarly, the point $\infty_{\rm N} \in U_{\rm S}$
parametrized by ${\bm x}_{\rm S}=0$ and ${\bm \q}_{\rm S} =0$
is infinitely separated for any observer in  $U_{\rm N}$.

In what follows, we will mostly work in the north chart and omit the subscript `N'
if no confusion may occur.
In the north chart, the null conditions \eqref{3.19} become
\begin{subequations}\label{3.35}
\bea
\bm \q^\dagger \O \bm \q + \ri \,( \bm x - \bm x^\dagger ) &=& 0 ~; \\
- \bm \q^{\rm T} \bm \q +\ri ( \s_2 \bm x +  \bm x^{\rm T} \s_2  )  &=&0~.
\label{3.32}
\eea
\end{subequations}
The reality conditions \eqref{3.20} turn into
\begin{subequations}\label{3.36}
\bea
 \overline{\bm \q}     &=& - \O \bm \q \s_2  ~;  \label{3.36a} \\
\overline{ \bm x}   &=&-\s_2 \bm x \s_2~.
\eea
\end{subequations}
It follows from eqs. \eqref{3.35} and \eqref{3.36} that
\bea
\bm x = x + \frac{\ri}{2}  \bm \q^\dagger\O \bm \q
= x + \frac{\ri}{4} \tr(\bm \q^\dagger\O \bm \q) {\mathbbm 1}_2 ~,
\qquad x = \vec{x} \cdot \vec{\s}~, \qquad \vec{x} \in {\mathbb R}^3~.
\label{3.37}
\eea
Thus the chart $U_{\rm N}$ may be identified with
a superspace ${\mathbb R}^{3|4n}$.


\subsection{Superconformal transformations}\label{Subsection3.5}

In the matrix realization \eqref{MR4.27},
any element  $\cL$ of the superconformal algebra
$\frak{osp}(2n|2,2)$ obeys the equations
\bea
\bm \cL^\dagger {\bm \X} +{\bm \X}\bm  \cL &=&0~, \qquad
\bm \cL^{\rm  sT} \bm \U +\bm \U \bm \cL =0~.
\eea
The general solution of these equations in the chosen parametrization is
\bea
\bm \cL = \left(
\begin{array}{c| c  c}
  u   & \bm \eta & \bm \e\\
 \hline
 -\bm \e^\dagger \O &    \hf \l {\mathbbm 1}_2 +\ri \vec{a}\cdot \vec{\s}
& \ri \vec{b}\cdot \vec{\s}  \\
-\bm \eta^\dagger \O  & \ri \vec{c}\cdot \vec{\s}  & - \hf \l {\mathbbm 1}_2 +\ri \vec{a}\cdot \vec{\s}
\end{array} \right)~, \qquad u \in {\frak so}^* (2n)~.
\label{3.39}
\eea
Here the bosonic parameters $\l$ and   $\vec{a}$, $\vec{b}$, $\vec c$ are
real, while the fermionic parameters obey the pseudo-Majorana condition
\bea
\overline{\bm \e} = -\O\bm \e \s_2~, \qquad
\overline{\bm \eta} = -\O\bm \eta \s_2~.
\label{3.40}
\eea
In what follows, we will use the condensed notation
\bea
{\frak a} := \vec{a} \cdot \vec{\s}~, \quad
{\frak b} := \vec{b} \cdot \vec{\s}~, \quad {\frak c} := \vec{c} \cdot \vec{\s}~,
\qquad \vec{a}, \, \vec{b}, \, \vec{c} \in {\mathbb R}^3
\label{3.43}
\eea
for the parameters in \eqref{3.39}.

Similar to the bosonic case, eq. \eqref{2.41},
the superconformal group acts on $S^{3|4n}$ by fractional
linear transformations.
In the infinitesimal case, the
superconformal transformation of $S^{3|4n}$
associated with $\bm \cL$, eq. \eqref{3.39}, is
\begin{subequations} \label{scxt}
\bea
\d \bm x &=&
{\frak b}
+ \l \bm x + \ri [
{\frak a}
, \bm x ]
+\ri \bm \e^\dagger \O \bm \q + \bm x
{\frak c}
\bm x
+\bm x \bm \eta^\dagger \O \bm \q~,   \label{scxta} \\
\d \bm \q &=& \bm \e  +\hf \l \bm \q + u \bm \q -\ri \bm \q
{\frak a}
+\ri \bm \eta \bm x +\bm \q \bm \eta^\dagger \O \bm \q
+\bm \q
{\frak c}
\bm x~.  \label{scxt-b}
\eea
\end{subequations}
Using these expressions, we can read off the superconformal transformation
of the  bosonic coordinates $x^a$ by representing
$x =\vec{x}\cdot \vec{\s} =  (x_\a{}^\b) $
in the form  $x= \bm x - \frac{\ri} {2} \bm \q^\dagger \O \bm \q$.

The isotropy group of the point $\infty_{\rm N} \in S^{3|4n}$
is generated by those
supermatrices \eqref{3.39} for which $\bm \eta =0$ and $\vec{c} =0$.
The most general element of the isotropy group of $\infty_{\rm N}$
is the product of a block-diagonal supermatrix
\bea
 \left(
\begin{array}{c| c  c}
  {\mathbbm 1}_{2n}   &  0& 0\\
 \hline
0 &  \re^{\hf \l} {\mathbbm 1}_2 &0 \\
0  & 0  &  \re^{-\hf \l} {\mathbbm 1}_2
\end{array} \right)
 \left(
\begin{array}{c| c  c}
{\frak U}   &  0& 0\\
 \hline
0 &   {\mathbbm 1}_2&~0 \\
0  & 0  &  {\mathbbm 1}_2
\end{array} \right)
 \left(
\begin{array}{c| c  c}
  {\mathbbm 1}_{2n}   &  0& 0\\
 \hline
0 &  {\frak R} &~0 \\
0  & 0  &  {\frak R}
\end{array} \right)
\label{block-diag}
\eea
with a super-translation
\bea
g ({\frak b} , \bm \e)= \left(
\begin{array}{c| c c}
  {\mathbbm 1}_{2n}   &~  0~& ~\bm \e\\
 \hline
-\bm \e^\dagger \O & ~ {\mathbbm 1}_2~&~ \ri \bm b  \\
0  &~ 0  ~& ~ {\mathbbm 1}_2
\end{array} \right)~.
\label{super-translation}
\eea
The parameters $\l$ and $\frak R$ in \eqref{block-diag} are the same as in \eqref{2.43},
and the matrix ${\frak U}$ is a group element of  $\sSO^* (2n)$,  see  \eqref{B.5}.
The fermionic parameter $\bm \e$ in \eqref{super-translation}
obeys the  pseudo-Majorana condition \eqref{3.40},
and the bosonic $ 2 \times 2$ matrix $\bm b$ has the form
\bea
\bm b = {\frak b}+ \frac{\ri}{2} \bm \e^\dagger\O \bm \e
={\frak b}+ \frac{\ri}{4} \tr(\bm \e^\dagger\O \bm \e) {\mathbbm 1}_2 ~,
\eea
with $\frak b $ being as in \eqref{3.43}.

All transformations \eqref{block-diag} also belong to the isotropy group
 of the point $\infty_{\rm S} \in S^{3|4n}$, which is the origin of
 the chart $U_{\rm N}$.  In addition, this group includes
all special conformal super-translations of the form
 \bea
\bm g ({\frak c} , \bm \eta)= \left(
\begin{array}{c| c  c}
  {\mathbbm 1}_{2n}   &~  \bm \eta~& ~0 \\
 \hline
0 & ~ {\mathbbm 1}_2~&~ 0  \\
 -\bm \eta^\dagger \O &~ \ri \bm c  ~& ~ {\mathbbm 1}_2
\end{array} \right)~.
\label{super-boost}
\eea
Here the fermionic parameter $\bm \eta$
obeys the  pseudo-Majorana condition \eqref{3.40},
and the bosonic $ 2 \times 2$ matrix $\bm c$ has the form
\bea
\bm c = {\frak c}+ \frac{\ri}{2} \bm \eta^\dagger\O \bm \eta
={\frak c}+ \frac{\ri}{4} \tr(\bm \eta^\dagger\O \bm \eta) {\mathbbm 1}_2 ~,
\eea
where $\frak c$ is defined by  \eqref{3.43}.
The supermatrices \eqref{block-diag}, \eqref{super-translation} and \eqref{super-boost}
generate the superconformal group $\sOSp(2n|2,2)$.
This statement is a version of the Harish-Chandra decomposition,
see, e.g., \cite{Knapp}.

The supermatrices \eqref{block-diag} with $\l=0$ and  \eqref{super-translation}
generate the isometry supergroup of a flat Euclidean superspace ${\mathbb E}^{3|4n}$.
As a supermanifold, this superspace may be identified with the north chart $U_{\rm N}$
of $S^{3|4n}$.
The action of the group elements  \eqref{block-diag} with $\l=0$ and  \eqref{super-translation}
on ${\mathbb E}^{3|4n}$ is induced by their action on $S^{3|2n}$.
In particular, the super-translation \eqref{super-translation}
acts on $S^{3|4n}$ by the rule
$\bm \cP \to \bm \cP' = \bm g({\frak b} , \bm \e) \bm \cP$,
with the two-plane $\bm \cP $
given by \eqref{3.34north}. The explicit form of this transformation is
\bea
x' =  x
+\frac{\ri}{2}  {\bm \e}^\dagger \O \bm \q
- \frac{\ri}{2}  {\bm \q}^\dagger \O \bm \e~,
\qquad {\bm \q}' = \bm \q + \bm \e~,
\eea
where we have used the transformation law
$\bm x' = \bm x  +\ri {\bm \e}^\dagger \O \bm \q + \frac{\ri}{2}  {\bm \e}^\dagger \O \bm \e$.

Let us consider the one-form\footnote{This one-form is a Euclidean 3D version
of the Volkov-Akulov supersymmetric one-form \cite{VA,AV}.}
\bea
e =  \rd x + \frac{\ri}{2} \rd \bm \q^\dagger \O \, \bm \q
- \frac{\ri}{2} \bm \q^\dagger \O \,\rd \bm \q
~, \quad
e_\a{}^\b = (\s^a)_\a{}^\b e_a~
~.~~~
\label{3.45}
\eea
The $\sSO^*(2n) $ transformations  \eqref{block-diag}  and
the super-translations  \eqref{super-translation} leave this one-form invariant.
Under the $\frak R$-transformations   \eqref{block-diag}, the one-form
changes as $e' = \frak R e{\frak R}^{-1}$.
As a result,
all transformations \eqref{block-diag} with $\l=0$ and  \eqref{super-translation}
leave invariant the metric
\bea
\rd s^2_{\rm flat} := e^a e_a~,
\label{flat_metric}
\eea
and therefore these transformations are indeed isometries of ${\mathbb E}^{3|4n}$.

\subsection{Superconformal metric}

Let us introduce a matrix two-point function on $S^{3|4n}$
\bea
\cE(1,2) := \cP^\dagger_1 \X \cP_2
= \Q^\dagger_1 \O_2 \Q_2+ {\bm h}^\dagger _1 {\bm h}_2 -{\mathbbm 1}_2~,
\label{3.42}
\eea
where $\cP$ is defined by \eqref{3.18}.
 Given a group element $g \in \sOSp(2n|2,2) $, it acts on $S^{3|4n}$ by the rule
 \bea
 g \left(
\begin{array}{c}
 \Q  \\  \hline
 \bm h \\
 {\mathbbm 1}_2
\end{array}
\right)
=
\left(
\begin{array}{c}
 \Q'  \\  \hline
  {\bm h}' \\
 {\mathbbm 1}_2
\end{array}
\right) \vf (g, \Q, \bm h)~, \qquad \vf (g, \Q, \bm h) \in \sGL (2, {\mathbb C})~.
\eea
This means that $\cE(1,2)$ transforms homogeneously,
\bea
\cE(1', 2') =  \big(\vf^\dagger (g, 1) \Big)^{-1}\cE(1,2)\,
\Big(\vf (g, 2) \Big)^{-1}~.
\eea
Associated with $\cE(1,2)$ is the two-point function  $\D (1,2) := \det \cE (1,2)$
with the superconformal transformation law
\bea
\D ( 1',2' ) =  \D (1,2) \Big(\overline{ \det \vf (g,1) }   \det \vf (g,2) \Big)^{-1} ~.
\eea

Let us choose $\Q_1= \Q$, $\bm h_1 = \bm h$ and $\Q_2 = \Q +\rd \Q$,
$\bm h_2 = \bm h + \rd \bm h$ in the definition \eqref{3.42}.
This gives the one-form
\bea\label{b1f}
\cE = \Q^\dagger \O \rd \Q + \bm h^\dagger \rd \bm h
= -\s_2 \Q^{\rm T} \rd \Q + \s_2 \bm h^{\rm T} \s_2 d\bm h~, \qquad
\cE^\dagger = - \cE
\eea
with the superconformal transformation
\bea
\cE' = (\vf^\dagger)^{-1} \cE \vf^{-1}~, \qquad \vf \equiv   \vf (g, \Q, \bm h) ~.
\eea

Introducing a super-interval
\bea
\rd s^2 := \frac{1}{4} \det \cE~,
\eea
it follows that it only scales under the superconformal transformations,
\bea
\rd s^2 ~\to ~ \rd s^2 | \det \vf |^{-2}~.
\eea
By construction, the super-interval is invariant under
the subgroup $\sOSp (2n|2) \times \sSU(2) \subset \sOSp(2n|2,2)$
which consists of those group elements which leave invariant
 the {\it non-null} two-plane
\bea
\left(
\begin{array}{c}
 0\\
\hline
 0 \\
 {\mathbbm 1}_2
\end{array}
\right) ~. \label{non-nulltwo-plane}
\eea

In the north chart, a direct calculation of $\cE$ gives the following expression:
\bea
\cE=  2 \ri ({\mathbbm 1}_2 +\ri {\bm x}^\dagger)^{-1} e \,
({\mathbbm 1}_2 -\ri \bm x)^{-1}~,
\label{3.54}
\eea
where $e= (e_\a{}^\b)$ is the rigid supersymmetric one-form
\eqref{3.45}.\footnote{This parametrization of the bosonic Cartan superform on $S^{3|4n}$ is similar
to a so-called GL-flat parametrization of the Cartan forms of the $\sOSp(1|2n,{\mathbb R})$ supergroup manifolds found in \cite{Plyushchay:2003gv}. More generally, the expression \eqref{3.54} is a natural extension of
those for the  Cartan forms on Hermitian symmetric spaces \cite{AKL}.}
As a result, the super-interval is
\bea
\rd s^2 = \frac{e^a e_a}{|\det   ({\mathbbm 1}_2 -\ri \bm x) |^2}~.
\label{3.55}
\eea
Switching off the Grassmann coordinates in \eqref{3.55} gives
a conformally covariant and $\sSO(4)$ invariant metric on $S^3$.
The supermetric \eqref{3.55} is a smooth
tensor field over $S^{3|4n}$.


\section{$\cN=2$ supersphere}

In this section we study the $n=1$ case. Its special feature is that
the $R$-symmetry subgroup of $\sOSp (2|2,2)$
is compact, $\sSO^*(2) \cong \sU(1)$.
For  all other values of $n>1$, the $R$-symmetry subgroup $\sSO^*(2n)$ of
the superconformal group $\sOSp (2n|2,2)$ is non-compact.
Since for $n=1$ the most general expression for $\O $
is $\pm \s_2$,
without loss of generality we  choose $\O = \s_2$.

It is useful to introduce new Grassmann coordinates,
$\bm \q_i{}^\a \to \hat{\bm \q}_i{}^\a
$, that have definite $\sU (1)_R$ charges.
They are defined as
\bea
\hat{\bm \q }
= (\hat{\bm \q}_i{}^\b )
\equiv
\left(
\begin{array}{c}
 \q^\b\\
 \bar \q^\b
 \end{array}
\right)
:=
T \bm \q
~, \qquad T = \frac{1}{\sqrt{2}}
\left(
\begin{array}{c c}
 1& ~\ri \\
 \ri  &~ 1
 \end{array}
\right) ~.
\label{4.1}
\eea
In this coordinate system, the super-metrics \eqref{3.30} become
\bea\label{sigma3}
\hat{\bm \X} = \left(
\begin{array}{c|l c}
 \hat \Omega  & ~0 & 0\\
 \hline
 0 & ~   0 & {\mathbbm 1}_2 \\
0 &~ {\mathbbm 1}_2 & 0
\end{array} \right)~,
\qquad
\hat{\bm \U} = \left(
\begin{array}{c|l c}
-\ri \tau   & ~0 & 0\\
 \hline
0 & ~   0 & \s_2 \\
0 &~ \s_2 & 0
\end{array} \right)~,
\eea
where  $\hat \O := T\O T^{-1} = -\s_3 $ and $\tau=\s_1$ with $\s_1$ and $\s_3$ being the first and third Pauli matrices carrying $SO(2)$ indices.
It is important to point out that $\hat \O$ is not
antisymmetric, unlike $\O$.
In the  coordinate system introduced, the null conditions \eqref{3.35} take the form
\begin{subequations}\label{4.2}
\bea
\hat{\bm \q}^\dagger \hat\Omega  \hat{\bm \q} + \ri \,( \bm x - \bm x^\dagger ) &=& 0 ~; \\
\hat{\bm \q}^{\rm T} \tau \hat{\bm \q} + \s_2 \bm x +  \bm x^{\rm T} \s_2    &=&0~.
\eea
\end{subequations}
The reality condition \eqref{3.36a} now reads
\bea
\overline{\q^\a} = \ve_{\a \b} \bar \q^\b\equiv \bar \q_{\a}~,
\qquad
\overline{ \bar \q{}^\a } = -\ve_{\a \b} \q{}^\b\equiv - \q_{\a}~.
\eea
To raise and lower two-component spinor indices,
we use antisymmetric matrices $\ve_{\a \b} = -\ve_{\b \a} $
and  $ \ve^{\a \b} = -\ve^{\b\a} $ normalized by
$\ve^{12}= - \ve_{12} =1$. The spinor indices are  lowered and raised
according to
\bea
\J^\a~ \to ~\J_\a = \ve_{\a \b} \J^\b~, \qquad
\J_\a~ \to ~\J^\a = \ve^{\a \b} \J_\b~.
\eea
Eq. \eqref{3.37} becomes
\bea
\bm x = x + \frac{\ri}{2} \q^\g \bar \q_\g  {\mathbbm 1}_2 ~,
\qquad x = \vec{x} \cdot \vec{\s}~, \qquad \vec{x} \in {\mathbb R}^3~.
\eea


\subsection{Superconformal transformations}

Here we specify  the main results of subsection \ref{Subsection3.5}
to the $n=1$ case  using the Grassmann coordinate basis introduced above.
The relations given in this  subsection are preparatory for our subsequent analysis
in the remainder of the section.

In the basis \eqref{4.1}, the  element \eqref{3.39} of the
superconformal algebra $\frak{osp}(2|2,2)$ takes the form:
\bea
\hat{\bm \cL} = \left(
\begin{array}{c| c  c}
 - \ri \vf \hat \Omega   &\hat{\bm \eta} & \hat{\bm \e}\\
 \hline
 -\hat{\bm \e}^\dagger \hat \Omega &    \hf \l {\mathbbm 1}_2 +\ri
{\frak a}
& \ri
{\frak b}
\\
-\hat{\bm \eta}^\dagger \hat\Omega  & \ri
{\frak c}
  & - \hf \l {\mathbbm 1}_2 +\ri
{\frak a}
\end{array} \right)~, \qquad \vf  \in {\mathbb R}~.
\eea
Here the bosonic parameters $\l$ and   $\frak a= \vec{a}\cdot \vec{\s}$,
$\frak b= \vec{b}\cdot \vec{\s}$, $\frak c= \vec{c}\cdot \vec{\s}$
are the same as in \eqref{3.39}. The fermionic $2\times 2$ matrix $\hat{\bm \e}$
has the structure
\bea
\hat{\bm \e }= (\hat{\bm \e}_i{}^\b )
\equiv
\left(
\begin{array}{c}
\e^\b\\
\bar \e^\b
\end{array}
\right)~, \qquad \overline{\e{}^\a} = \bar \e_{ \a} = \ve_{\ab}\bar \e{}^\b~,
\qquad \overline{\bar \e{}^\a} = - \e_{ \a} ~,
 \eea
and similar for $\hat{\bm \eta}$.
The parameter $\vf$ describes a $\sU(1)_R$ transformation.
The $\sU(1)_R$ charge of $\q^\a$ is $+1$.
As follows from \eqref{scxt}, the most general
 infinitesimal superconformal transformation in the north chart  of $S^{3|4}$ is
\begin{subequations}\label{4.9}
\bea
\d \bm x &=& \frak b
+ \l \bm x + \ri [  \frak a  , \bm x ]
+\ri \hat{\bm \e}^\dagger \hat\Omega \hat{\bm \q} + \bm x
\frak c
\bm x
+\bm x \hat{\bm \eta}^\dagger \hat\Omega \hat{\bm \q}~, \\
\d \hat{\bm \q} &=& \hat{\bm \e}  +\hf \l \hat{\bm \q}
- \ri \vf \hat \O \hat{\bm \q} -\ri \hat{\bm \q}
\frak a
+\ri \hat{\bm \eta} {\bm x} +  \hat{\bm \q}  \hat{\bm \eta}^\dagger \hat\Omega \hat{\bm \q}
+\hat{\bm \q}
\frak c
 \bm x~.
\eea
\end{subequations}

The super-translation \eqref{super-translation} takes the form
\bea\label{bsuper}
\hat{\bm g} (\frak{b} , \hat{\bm \e}) = \left(
\begin{array}{c| c  c}
{\mathbbm 1}_2   &0 & \hat{\bm  \e}\\
 \hline
-\hat{\bm \e}^\dagger \hat\Omega &   {\mathbbm 1}_2 & \ri \bm b \\
0  & 0  &  {\mathbbm 1}_2
\end{array} \right)~, \qquad
\bm b=
\frak b
+ \frac{\ri}{2} \e^\g \bar \e_\g  {\mathbbm 1}_2 ~.
\eea
In the north chart  of $S^{3|4}$, this group element acts as follows
\bea
\hat{\bm  g} (\frak{b} , \hat{\bm \e})
\left(
\begin{array}{c}
 \hat{\bm \q} \\  \hline
  \ri {\bm x} \\
 {\mathbbm 1}_2
\end{array}
\right) = \left(
\begin{array}{c}
 \hat{\bm \q}{}' \\  \hline
  \ri {\bm x}' \\
 {\mathbbm 1}_2
\end{array}
\right) ~,
\label{4.11}
\eea
where
\bea
\bm x' = \bm x +
\bm b
+ \ri \hat{\bm \e}^\dagger \hat\Omega \hat{\bm \q} ~,
\qquad \hat{\bm \q}' = \hat{\bm \q} + \hat{\bm \e}~.
\label{4.12}
\eea
In terms of the coordinates $x_\a{}^\b$ and $\q_\a$, this transformation law
reads\footnote{The symmetrization of two spinor indices in \eqref{4.13} includes a factor of 1/2.}
\bea
x'_\a{}^\b = x_\a{}^\b +b_\a{}^\b - \ri \e_{(\a} \bar \q^{\b)}
- \ri \bar \e_{(\a} \q^{\b)} ~, \qquad
\q'_\a = \q_\a +\e_\a~.
\label{4.13}
\eea
The supersymmetric Cartan form \eqref{3.45} takes the form
\bea
e_\a{}^\b = \rd x_\a{}^\b +\ri \bar \q_{(\a} \rd \q^{\b)}
+\ri \q_{(\a} \rd \bar \q^{\b)}~, \qquad
e_\a{}^\b = (\s^a)_\a{}^\b e_a~.
\label{4.14}
\eea


\subsection{Chiral subspace}\label{cs}
Let us now introduce a complex three-vector variable\footnote{In the remainder of this section,
we often make use of  row vectors
$\q = (\q^\a)$ and $ \bar \q = (\bar \q^\a)$ and column vectors
$\tilde{\q} = (\q_\a)$ and $ \tilde{\bar \q} = (\bar \q_\a )$.}
$y^a$ defined by
\bea
y = \bm x + \ri \tilde{\q} \bar \q = \vec{y} \cdot \vec{\s}\quad \Longleftrightarrow
\quad y_\a{}^\b = x_\a{}^\b +\ri \q_{(\a} \bar \q^{\b)}~.
\label{4.15}
\eea
In accordance with
\eqref{4.13}, the transformation law of $y$ is
\bea
y'_\a{}^\b = y_\a{}^\b +b_\a{}^\b -2\ri \bar \e_{(\a} \q^{\b)}
+\ri \bar \e_{(\a} \e^{\b)} ~.
\eea
We see that the {\it chiral} variables $y^a$ and $\q^\a$ form a closed subset
under the super-transformations.

It is nontrivial that the chiral variables also form
a closed subset under the superconformal transformations.
Indeed, the infinitesimal superconformal transformation \eqref{4.9}
may be used to show that the chiral variables vary as follows:
\begin{subequations} \label{4.16}
\bea
\d y &=& \frak b
+ \l y + \ri [
\frak a, y ]
+\ri \tilde{\q} \e -\ri \tilde{\bar \e} \q
+ y \frak c
y
-(y\tilde{\bar \eta} )\q - \tilde{\q} (\bar \eta y)
~, \\
\d  \q &=& \e  +\hf \l \q
+ \ri \vf  \q -\ri \q \frak a
+\ri \eta y +  (\q  \tilde{\bar \eta}) \q
+ \q \frak c y~.
\eea
\end{subequations}

The above property allows us to give an alternative
definition  of the
3D $\cN=2$ superconformal group that is analogous to the one
used in \cite{BK} in  the 4D $\cN=1$ super-Poincar\'e case.
We introduce a complex superspace
${\mathbb C}^{3|2}$ parametrized by bosonic $y$ and fermionic $\q^\a$
variables. Embedded into ${\mathbb C}^{3|2}$ is a real superspace
${\mathbb R}^{3|4}$ with coordinates $z^A = (x^a, \q^\a, \bar \q_\a)$,
with $\bar \q_\a:= \overline{\q^\a}$,
which is defined by
\bea
y^a - \bar y^a = 2\ri \cH^a~, \qquad \cH^a := \hf {\q} \s^a \tilde{\bar \q}
=\hf \q^\a (\s^a)_\a{}^\b \bar \q_\b ~.
\label{4.17}
\eea
An infinitesimal holomorphic transformation on ${\mathbb C}^{3|2}$,
\bea
\d y^a = \x^a (y, \q) ~, \qquad \d \q^\a = \x^\a  (y, \q) ~,
\eea
is said to be superconformal if it preserves the real surface \eqref{4.17};
that is,
\bea
\x^a - \bar \x^a = \ri (\s^a)_\a{}^\b \Big( \x^\a \bar \q_\b + \q^\a \bar \x_\b \Big)~,
\eea
where $ \bar \x_\a (\bar y ,  \bar \q) := \overline{ \x^\a  (y, \q) }$.
It is an instructive exercise to show that the most general solution of this equation
is given by \eqref{4.16}.

\subsection{Complexified supersphere}
In accordance with \eqref{4.16}, the superconformal group acts by holomorphic
transformations on the chiral variables $\z_{\rm N}  = (y_{\rm N}{}^a  , \q_{\rm N}{}^\a)$
defined in the north chart $U_{\rm N}$ of $S^{3|4}$.
We can also introduce chiral variables $\z_{\rm S}  = (y_{\rm S}{}^a  , \q_{\rm S}{}^\a)$
defined in the south chart $U_{\rm S}$ of $S^{3|4}$, by
extending  the definition \eqref{4.15} to the south chart. It is natural to wonder
whether the concept of chirality is just a local structure defined within a coordinate chart
or if  it is globally defined on $S^{3|4}$.

In the overlap of the north and south charts, $U_{\rm N} \bigcap U_{\rm S} $,
we derive the transition functions:
\bea
y_{\rm S} &=& - y_{\rm N}{}^{-1}~, \qquad
\q_{\rm S }{}^\a = -\ri \,\q_{\rm N}{}^\b  (y_{\rm N}{}^{-1})_\b{}^\a~.
\label{4.21}
\eea
This result shows that chirality is globally defined on $S^{3|4}$.

It is natural to  introduce a complexified or chiral supersphere, ${\mathbb C}S^{3|2}$.
It is  defined to be a complex supermanifold which  may be
covered by two charts  $W_{\rm N} $ and $W_{\rm S} $,
${\mathbb C}S^{3|2}= W_{\rm N} \bigcup W_{\rm S} $,
 such that   the following properties hold:
(i) each chart is diffeomorphic to complex superspace ${\mathbb C}^{3|2}$
parametrized by independent complex coordinates $\z = (y^a, \q^\a)$;
and (ii) in the overlap of the charts,  $W_{\rm N} \bigcap W_{\rm S} $,
the local coordinates are related to each other by the transition
functions \eqref{4.21}.
The superconformal group naturally acts on ${\mathbb C}S^{3|2}$
by holomorphic transformations  \eqref{4.16}.
The bosonic body of ${\mathbb C}S^{3|2}$ is a complexified three-sphere
that may be identified
with the tangent bundle $T S^3$ of the three-sphere.\footnote{As is known, the
 complexified three-sphere may be realized as a quadric in ${\mathbb C}^4$
 defined by $\vec{Z} \cdot \vec{Z} = 1$, with $\vec{Z}=\vec{X} +\ri \vec{Y} \in {\mathbb C}^4$
 and $\vec{X}, \vec{Y} \in {\mathbb R}^4$.}


\subsection{Superconformal inversion}

Super-inversion is a discrete transformation $I_k : S^{3|4} \to S^{3|4}$
defined by
\bea
\q' &=& \bar \k \bar \q (y^\dagger)^{-1}
~, \qquad
y' = |\k|^2 (y^\dagger)^{-1} ~,
\eea
for some non-zero parameter $\k$.
This parameter may always be chosen to be equal to any given nonzero
complex number by combing $I_\k$ with a scale and $\sU(1)_R$ transformation.
One may check that $(I_\k )^2= \rm{id}$.
The super-inversion respects the defining equation of the chiral subspace,
\bea
\ri (y' -y'^\dagger )_\a{}^\b = 2\bar \q'_{(\a} \q'^{\b)}~.
\eea

It is an instructive exercise to show that
the super-inversion is a discrete superconformal transformation
in the sense that it only rescales
the flat supermetric \eqref{flat_metric},
\bea
\tr (e')^2 = \frac{|\k |^4}{y^2 \bar y^2} \tr (e^2)~,
\eea
with the supersymmetric Cartan form given by \eqref{4.14}.
If one considers a composite transformation
$I_\k  \, \hat{\bm  g} (\frak{b} , \hat{\bm \e}) \,I_\k$,
with $\hat{\bm  g} (\frak{b} , \hat{\bm \e})$ being the super-translation \eqref{bsuper},
the resulting transformation is a special conformal super-translation.

The above properties are analogous to those possessed by a super-inversion
in the case of 4D $\cN=1$ superconformal symmetry \cite{GGRS,BK}.


\section{Supercoset realizations of  $\mathbb E^{3|4n}$ and $S^{3|4n}$ }

In this section we give several supercoset realizations for  $S^{3|4n}$ and
flat Euclidean superspace $\mathbb E^{3|4n}$.

\subsection{The super-translation subalgebra of $\frak{osp}(2n|2,2)$ and $\mathbb E^{3|4n}$}

The Euclidean counterpart of the $D=3$, $\mathcal N=2n$ super-Poincar\'e algebra is obtained from \eqref{D3osp2n22} by projecting the supersymmetry generators ${\mathcal Q}^i_{\hat\alpha}$ as follows
\be\label{P04Q}
\tilde Q^i={\mathcal Q}^i{\mathbb P}_{04}\,,
\ee
where
\be\label{P04}
{\mathbb P}_{04}=\frac 12 ({\mathbbm 1}+\gamma^0\gamma^4)\,,
\qquad {\mathbb P}_{04}{\mathbb P}_{04}={\mathbb P}_{04}\,.
\ee
The supercharges \eqref{P04Q}, whose number is half the number of ${\mathcal Q}^i$, are transformed under the fundamental representation of the group $\sSU(2)$ of rotations in $D=3$ labeled by the index $\alpha=1,2$. They generate a superalgebra which is obtained from \eqref{D3osp2n22} by multiplying its left and right hand sides by the projectors ${\mathbb P}_{04}$, taking into account the order of the spinor indices. Due to the
anti-commutation properties of the gamma-matrices, the terms on the right hand side of  \eqref{D3osp2n22} which survive this projection have the following form
\be\label{3Dspoincare}
\{\tilde Q^i,\tilde Q^j\}=2\delta^{ij}\, \sigma^a\, P_a\,, \qquad [P_a,P_b]=0\,.
\ee
Due to the chosen realization of the gamma-matrices,
$$\ri\sigma^a=({\mathbbm 1}-{\mathbb P}_{04})\gamma^a\gamma^0\,{\mathbb P}_{04}=({\mathbbm 1}-{\mathbb P}_{04})\gamma^a\gamma^4\,{\mathbb P}_{04}$$
 can be associated with the Pauli matrices and $P_a=\ri (L_{a0}+L_{a4})$ is the generator of the translations in $3d$ flat space.  The projections  $({\mathbbm 1}-{\mathbb P}_{04})\gamma^{ab} {\mathbb P}_{04}$ and $({\mathbbm 1}-{\mathbb P}_{04})\gamma^{04}{\mathbb P}_{04}$ vanish due to the commutation properties of the gamma-matrices.

\if{}
Explicitly the projection goes as follows
\bea\label{PP04}
\frac 14(\delta^{\hat\alpha'}_{\hat\alpha}+\gamma^{04\hat\alpha'}{}_{\hat\alpha})\gamma^{\hat a\hat b}_{\hat\alpha'\hat\beta'}(\delta^{\hat\beta'}_{\hat\beta}+\gamma^{04\hat\beta'}{}_{\hat\beta})
=\frac 14(C_{\hat\alpha\hat\alpha'}-\gamma^{04}_{\hat\alpha'\hat\alpha})(\gamma^{\hat a\hat b})^{\hat\alpha'}{}_{\hat\beta'}(\delta^{\hat\beta'}_{\hat\beta}+\gamma^{04\hat\beta'}{}_{\hat\beta})
\nonumber\\
=\frac 14C_{\hat\alpha\hat\gamma}(\delta^{\hat\gamma}_{\hat\alpha'}-\gamma^{04\hat\gamma}{}_{\hat\alpha'})(\gamma^{\hat a\hat b})^{\hat\alpha'}{}_{\hat\beta'}(\delta^{\hat\beta'}_{\hat\beta}+\gamma^{04\hat\beta'}{}_{\hat\beta})
=C(\mathbb{I}-{\mathbb P}_{04})\gamma^{\hat a\hat b}{\mathbb P}_{04},
\eea
\fi

The $\sSU(2)\cong \sSO(3)/{\mathbb Z}_2$ group, under which $\tilde Q^i$ and $P_a$ transform in  the spinor and the vector representations, respectively, is generated by the operators $L_{ab}$, while $\sSO^*(2n)$ generated by $T^{ij}$ becomes the group of ``external" $R$--symmetries of this superalgebra.

In the diagonal matrix realization\footnote{This realization is obtained from that of \eqref{gamma} by applying the similarity transformation \eqref{2.22}.} of the projector ${\mathbb P}_{04}$,
\be\label{P040}
{\mathbb P}_{04}=\left(\begin{tabular}{ c c}
 0& 0 \\
0 &  $\mathbbm 1$\\
\end{tabular}
\right)\,,
\ee
the
elements of the $3D$ super--translation group associated with the Euclidean superspace $\mathbb{E}^{3|4n}$ are similar to
\eqref{super-translation},
\bea\label{cosets3}
\mathbb{E}^{3|4n}(x,\theta) = \left(
\begin{array}{c| c  c}
  {\mathbbm 1}_{2n}   &~  0~& ~\bm \theta\\
 \hline
-\bm \theta^\dagger \O & ~ {\mathbbm 1}_2~&~ \ri \bm x  \\
0  &~ 0  ~& ~ {\mathbbm 1}_2
\end{array} \right)~ ,
\eea
where $\bm x=x^a\sigma^a + \frac{\ri}{2} \bm \theta^\dagger\O \bm \theta$ and the spinors $\bm\theta$ satisfy the symplectic--Majorana reality condition \eqref{3.36a} which follows from the reality condition \eqref{sM} for the projected supercharges \eqref{P04Q}.

Note that the right column in \eqref{cosets3} is nothing but the two-plane \eqref{3.34north} which
describes a point in  the north chart of $S^{3|4n}$.

The superspace ${\mathbb E}^{3|4n}$ defined in \eqref{cosets3} can be regarded as a {\it local}
supercoset of the superconformal group, namely
\be\label{ssC}
{\mathbb E}^{3|4n}
\subset S^{3|4n}
=\frac{\sOSp(2n|2,2)}{\sSO^*(2n)\times \sSU(2) \rtimes SK}~,
\ee
where $SK$ stands for the dilatation, conformal boosts and superconformal transformations.
In other words,
the stability group $\mathbb H$ of this coset
is formed by the product of the matrices \eqref{block-diag} and \eqref{super-boost}.
We recall that $\mathbb H$ is the isotropy group of
 the point $\infty_{\rm N} \in S^{3|4n}$
(see subsection \ref{Subsection3.5}) and the superspace ${\mathbb E}^{3|4n}$ can be identified with
\bea
{\mathbb E}^{3|4n} = S^{3|4n} \setminus \{ \infty_{\rm N} \}~.
\eea

The superconformal group generated by \eqref{3.39} acts on the superspace ${\mathbb E}^{3|4n}$ coset element \eqref{cosets3} as follows
\be\label{scaction}
{\mathbb E}'(x',\theta')=\re^{\bm \cL}\,{\mathbb E} (x,\theta ){\mathbb H}^{-1}(x,\theta)~,
\ee
where ${\mathbb H}^{-1}(x,\theta)$ is the compensating transformation from  the stability group,
which is required in order to bring the transformed coset element to a form similar to \eqref{cosets3}. One can check that the transformation \eqref{scaction} with infinitesimal parameters generates the superconformal transformations of $x$ and $\theta$ given in \eqref{scxt}.

The special conformal super-translations \eqref{super-boost}
do not generate a  well defined action on the flat superspace
${\mathbb E}^{3|4n} $ if the body of the special conformal parameter $c^a$ in
$ \frak c = \vec{c} \cdot \vec{\s} $ is non-zero. In this case some point $(x_0, \q_0)$
from ${\mathbb E}^{3|4n} $ is mapped to the infinitely separated point,
$ \infty_{\rm N} $, which means that ${\mathbb H}^{-1}(x_0,\theta_0)$ is not defined.
By construction, all elements of the superconformal group generate well defined transformations
on the supersphere $S^{3|4n}$.

As we discussed in Section \ref{cs}, in the $n=1$ case in which
$\sSO^*(2)= \sSO(2)$, there is a chiral subspace
which transforms into itself under the super-translation and the infinitesimal superconformal
transformations. In the generic $n>1$  case, there is no chiral subspace which would transform
into itself under $\sSO^*(2n)$, since the $\sSO^*(2n)$ matrices (with $n>1$)
do not commute with the symplectic form $\Omega$. The same conclusion also follows from the fact that
the defining representation of  $\sSO^*(2n)$ is irreducible for $n>1$.


\subsection{The $\sOSp(2n|2)\times \sSU(2)$ subalgebra of $\frak{osp}(2n|2,2)$ and $S^{3|4n}$}

We recall that
the super-interval \eqref{3.55} is invariant under  a
subgroup $\sOSp (2n|2)\times \sSU(2) $ of
the superconformal group $\sOSp (2n|2,2)$.
In the matrix realization \eqref{slashX} of $\sOSp (2n|2,2)$,
this subgroup consists of those group elements which leave invariant
 the two-plane \eqref{non-nulltwo-plane}.
The bosonic subgroup of the supergroup $\sOSp(2n|2)$ is $\sSO^*(2n)\times \sSp(2)$, where $\sSp(2)\cong \sSU(2)$.

To get an $\sOSp(2n|2)\times \sSU(2)$ sub-superalgebra of the $\sOSp(2n|2,2)$ superconformal algebra \eqref{D3osp2n22}, one may single out half of the supergenerators ${\mathcal Q}^i$ using the projector
\be\label{P0}
 {\mathbb P}_0=\frac 12({\mathbbm 1}+ \gamma_0)\,,\qquad {\mathbb P}_0{\mathbb P}_0={\mathbb P}_0\,,
\ee
as follows
\be\label{su2n}
 Q_{\hat\alpha}^i=({\mathcal Q}^i\,{\mathbb P}_0)_{\hat\alpha}=(Q^i_\alpha,0)\,, \qquad \alpha=1,2\,,
\ee
where the index $\a$ corresponds to
the $\sSU(2)\cong \sSp(2)$ subgroup of $\sOSp(2n|2)$.

The spinors $Q^i_\alpha$ satisfy the symplectic-Majorana condition
\be\label{SM1}
Q^{i\alpha}=-\ri\varepsilon^{\alpha\beta} \Omega^{ij}\bar Q^{j}_\beta \equiv -\ri\Omega^{ij}\bar Q^{j\alpha}
=-\ri\Omega^{ij}\,( Q_{\alpha}^ j)^*\,.
\ee

Multiplying  both sides of \eqref{D3osp2n22}  by ${\mathbb P}_0$
and taking into account the order of the indices we get
\be\label{su2n1}
\{{ Q}^i_\alpha,{Q}^j_\beta\}=\delta^{ij}(-\sigma_{\alpha\beta}^{ab}\,L_{ab}+2\ri\sigma^a_{\alpha\beta}\,L_{a4})+ \epsilon_{\alpha\beta} T^{ij}\,,
\ee
where
\be\label{Pauli}
\sigma^a\equiv \ri{\mathbb P}_0\,\gamma^a\gamma^4\,{\mathbb P}_0,  \qquad \sigma^{ab} \equiv -{\mathbb P}_0\,\gamma^{ab}{\mathbb P}_0
\ee
can be associated with the Pauli matrices and $\epsilon=\ri\sigma_2={\mathbb P}_0\,C\,{\mathbb P}_0$,
while ${\mathbb P}_0\gamma^{\hat a 0}{\mathbb P}_0=0$, since $\gamma^a$ and $\gamma^4$
anticommute with $\gamma^0$ inside ${\mathbb P}_0$.

Furthermore, using the identity $\sigma^{ab}=\ri \varepsilon^{abc}\sigma_c$, we may rewrite \eqref{su2n1} as follows
\be\label{su2n2}
\{{ Q}^i_\alpha,{Q}^j_\beta\}=\delta^{ij}\,\sigma_{\alpha\beta}^{a}\,M_a+ \epsilon_{\alpha\beta} T^{ij}\,,
\ee
where
\be\label{Mm}
M_a=\ri(L_{a4}-\frac 12\varepsilon_{abc}\,L^{bc})
\ee
generate the $\sSU(2)$ algebra
\be\label{su2}
[M_a,M_b]=2\ri\varepsilon_{abc}\,M_c\,.
\ee
We see that the generators $\tilde M_a=\ri(\frac 12\varepsilon_{abc}\,L^{bc}+L_{a4})$ of another $\sSU(2)$ subalgebra  of $\sOSp(2n|2,2)$ do not appear in the right hand side of \eqref{su2n2} and thus commute with those of the $\sOSp(2n|2)$.

In the $n=1$ case, the superalgebra isomorphism
$\frak{osp}(2|2)\cong \frak{su}(2|1)$ holds.
Introducing
the complex conjugate supercharges
\be\label{QbarQ}
Q_\alpha=\frac 1{\sqrt 2}(Q^1_\alpha+\ri Q^2_\alpha)\,,
\qquad \bar Q_\alpha=-\frac 1{\sqrt 2}(Q^1_\alpha-\ri Q^2_\alpha)\,, \qquad (Q^\alpha)^*=\bar Q_\alpha\,,
\ee
the anti-commutation relations take the form
\be\label{su21}
\{Q_\alpha,\bar Q_\beta\}=\sigma_{\alpha\beta}^m\,M_m+\epsilon_{\alpha\beta} R\,, \qquad \{Q_\alpha, Q_\beta\}=0\,,
\ee
where $R$ is the $\sU(1)$ $R$-symmetry generator.

In accordance with the above consideration,
 every element $M \in \frak{osp}(2n|2)$ is
singled out from some element of the superconformal algebra,
$\mathcal M \in \frak{osp}(2n|2,2)$,
by multiplying the latter (from the left and from the right) with the projector
\be
{\mathcal P}_0=\left(\begin{tabular}{ c c }
 ${ \mathbbm 1}_{2n}$ & 0  \\
  0 &${\mathbb P}_0$ \\
\end{tabular}
\right),
\ee
namely
\be\label{MM}
M={\mathcal P}_0{\mathcal M}{\mathcal P}_0\,,
\ee
where ${ \mathbbm 1}_{2n}$ is the unit matrix acting on the $\sSO^*(2n)$ indices.

The supersphere can be identified
with the coset superspace
\bea
{\mathbb S}^{3|4n}=\frac{\sOSp(2n|2)}{\sSO^*(2n)}~,
\eea
which is formed by the equivalence classes
\be\label{SS3}
\re^M \sim \re^M h~, \qquad h\in  \sSO^*(2n)~,
\ee
where $M \in \frak{osp}(2n|2)$ is given by  \eqref{MM}.

In the gamma-matrix realization \eqref{gamma} in which
\be\label{p00}
{\mathbb P}_0=\left(\begin{tabular}{ c c  }
$\mathbbm{1}_{2}$  & 0 \\
0 & 0 \\
\end{tabular}
\right),
\ee
the algebra--valued element
\eqref{MM} associated with the ${\mathbb S}^{3|4n}$
coset generators of the supergroup $\sOSp(2n|2)$ is
\be
\label{cosets33}
{\mathbf s}^{3|4n}=\left(\begin{tabular}{ c c c}
0  & $\Theta$ &  0\\
 $ -\Theta^\dagger\Omega$ & $\ri \mathbf{x}
 $ &0 \\
 0 & 0 & 0\\
\end{tabular}
\right)\,,
\ee
where $\Theta$ are subject to the symplectic--Majorana condition \eqref{3.20} and
${\mathbf x}= ({\mathbf x}_\alpha{}^{\beta})$ is a traceless Hermitian  matrix.
We see that the rank of \eqref{cosets33} reduces to $2n+2$.

The supersphere $S^{3|4n}$ parametrized by \eqref{cosets33} can also be regarded as a supercoset of the conformal group $\sOSp(2n|2,2)$ in its realization defined in \eqref{slashX}, which is \emph{different} from \eqref{ssC}. The relevant supercoset is
\be\label{ssCS}
{\mathbb S}^{3|4n}=
\Big\{ \re^{{\mathbf s}^{3|4n}} \Big\}
=\frac{\sOSp(2n|2,2)}{\sSO^*(2n)\times \sSU(2) \rtimes SK}~,
\ee
where, as in \eqref{ssC}, $SK$ stands for the dilatation, conformal boosts and superconformal transformations. The stability  group $\hat {\mathbb H}=\sSO^*(2n)\times \sSU(2) \rtimes SK$ of this coset
is formed by the product of the matrices \eqref{block-diag} and \eqref{super-boost}
(as in \eqref{ssC} but) subject to the similarity transformation with the inverse matrix of \eqref{Sigma4.27},
namely
\be\label{SHS}
\hat{\mathbb H}={\bm \S}^{-1}{\mathbb H}{\bm \S}\,.
\ee
The superconformal transformation of the supercoset \eqref{ssCS} is
\be\label{scactionS}
{\mathbb S}({\mathbf x}',\Theta')=({\bm \S}^{-1}\re^{\bm \cL}{\bm \S})\,{\mathbb S} (\mathbf x,\Theta )\hat{\mathbb H}^{-1}(\mathbf x,\Theta)\,,
\ee
where $\bm{\cL}$ is the same as in  \eqref{3.39}.

The supercoset element associated with \eqref{cosets33}
parametrizing the points of the supersphere $S^{3|4n}$ can be given in the form
\be\label{cosetS}
{\mathbb { S}}=\left(\begin{tabular}{ c c}
$\mathbf M$ & $\Theta $ \\
$- {\bm h}\Theta^\dagger\Omega \mathbf M^{-1}$  & $ \bm {h}$ \\
\end{tabular}
\right)=\left(\begin{tabular}{ c c}
$\mathbf M$ & $\Theta $ \\
$\sigma_2({\bm h}^{-1})^{\rm  T}\Theta^{\rm T}\mathbf M$  & $ \bm {h}$ \\
\end{tabular}
\right)~,
\ee
where $\bm h$ satisfies the constraints \eqref{3.19} and
\eqref{3.20}, while $\bf M$ is defined by
\bea
\mathbf M :=({\mathbbm 1}_{2n}-\Theta\Theta^\dagger\Omega)^{\frac 12}
=(\mathbbm 1_{2n}+\Theta\sigma_2\Theta^{\rm T})^{\frac 12}
=\Omega \mathbf M^{\dagger}\Omega ~,
\eea
such that $(\mathbf M^2)_i{}^{j}=\delta_i^j+\ri\Theta_i^\alpha\Theta_\alpha^j$.
The right column of \eqref{cosetS}
involves the same matrix blocks $\Q$ and $\bm h$
which constitute the null two-plane \eqref{3.18}.
The inverse of $\mathbb S$ is
\be
{\mathbb { S}^{-1}}=\left(\begin{tabular}{ c c}
$\mathbf M$ & $-\mathbf M^{-1 }\Theta{\bm h}^{\dagger} $ \\
$ \Theta^\dagger\Omega$  & $ {\bm h}^{\dagger}$ \\
\end{tabular}
\right) =\left(\begin{tabular}{ c c}
$\mathbf M$ & $-\mathbf M \Theta\bm h^{-1} $ \\
$-\sigma_2 \Theta^{\rm T}$  & $ {\bm h}^{\dagger}$ \\
\end{tabular}
\right) ~.
\ee

For completeness, here we  give the most general element of
$\sOSp(2n|2)$:
\bea
g
=
\left(
\begin{array}{c|c}
\bf M  \frak U ~&~ \Q \\
 \hline
\s_2 ({\bm h}^{-1})^{\rm T} \Q^{\rm T} \bf M \frak U ~&  ~  \bm h
\end{array}
\right)~~, \qquad \frak U \in \sSO^*(2n)~.
\label{6.32}
\eea
The coset representative \eqref{cosetS}
is obtained from \eqref{6.32} by setting $\frak U = {\mathbbm 1}_{2n}$.

The Cartan form describing the  geometry of $S^{3|4n}$
in this realization is
\bea\label{S34ncartan}
{\mathbf { S}^{-1}}\rd{\mathbf { S}}&=&\left(\begin{tabular}{ c c}
${\mathbf M}\rd{\mathbf M}+\mathbf M^{-1 }\Theta{\bm h}^{\dagger}\rd({\bm h}\Theta^\dagger\Omega
\mathbf M^{-1})$ & ${\mathbf M} \rd\Theta -\mathbf M^{-1 }\Theta{\bm h}^{\dagger} \rd {\bm h}$ \\
$ \Theta^\dagger\Omega \rd {\mathbf M}-{\bm h}^{\dagger}\rd({\bm h}\Theta^\dagger\Omega \mathbf M^{-1})$
& $ {\bm h^\dagger \rd \bm h}+\Theta^\dagger \Omega \rd\Theta$ \\
\end{tabular}
\right)\nonumber\\
&=&\left(\begin{tabular}{ c c}
${\mathbf M}\rd{\mathbf M}-\mathbf M\Theta{\bm h}^{-1}\rd({\sigma_2
({\bm h}^{-1})^{\rm T}\Theta^{\rm T}\mathbf M})$
& ${\mathbf M}(\rd\Theta -\Theta{\bm h}^{-1}\rd {\bm h})$ \\
$ -\sigma_2\Theta^{\rm T} \rd {\mathbf M}+{\bm h}^{\dagger}\rd(\sigma_2
({\bm h}^{-1})^{\rm  T}\Theta^{\rm T}\mathbf M)$
& $ {\bm h^\dagger \rd \bm h}+\Theta^\dagger \Omega \rd\Theta$ \\
\end{tabular}
\right)~.~~~
\eea
This Cartan form completes the  $S^{3|4n}$ supervielbein derived in \eqref{b1f} with its fermionic counterpart $\cE^{\rm fer}$
and the $\sSO^*(2n)$-connection $\omega^{\frak{so}^*(2n)}$:
\bea
\cE^{\rm fer}={\mathbf M}(\rd\Theta -\Theta{\bm h}^{-1}\rd {\bm h})~, \quad
\omega^{\frak{so}^*(2n)}={\mathbf M}\rd{\mathbf M}
+\mathbf M^{-1 }\Theta{\bm h}^{\dagger}\rd({\mathbf h}\Theta^\dagger\Omega \mathbf M^{-1})~.~~~
\eea

In the $n=1$ case,  in which $\sSO^*(2)=\sSO(2)\cong \sU(1)$,
 the $\sOSp(2|2)$ supergroup is isomorphic to $\sSU(2|1)$. To reduce the $\sOSp(2|2)$ superalgebra valued element \eqref{cosets33} to a corresponding $\sSU(2|1)$ superalgebra element $\mathbf u$ it is convenient to use the projector
 ${\mathbb P}_\Omega=\frac 12({\mathbbm 1}+\Omega)$, then
\be\label{u}
\mathbf u=\mathbb P_\Omega\,\mathbf s^{3|4}\,\mathbb P_\Omega=\left(\begin{tabular}{ c c}
0  & $\theta$ \\
 $ -\bar\theta$ & $\ri\mathbf{x}
 $ \\
\end{tabular}
\right),
\ee
where $\theta=\mathbb P_\Omega\,\Theta$ and $\bar\theta=\Omega^\dagger\, \mathbb P_\Omega=(\theta)^\dagger$. Note that in the realization in which $\Omega=-\sigma_3$ (see \eqref{sigma3}),
the Grassmann variables $\theta^\alpha$ and $\bar\theta_\alpha$ transform under the complex conjugate one-dimensional representations of $\sU(1)$.
This just reflects the fact that the rank of \eqref{u} is 3, i.e. the same as  of the
$\sSU(2|1)$ superalgebra valued matrix generating an $\frac{\sSU(2|1)}{\sU(1)}$ coset element.


\subsection{$S^{3|4}$ in the matrix realization of $\sSU(2|1)$}

For completeness, let us now consider the description of the supersphere  $S^{3|4}$ as the supercoset\footnote{For the construction of quantum mechanical models on different
cosets of $\sSU(2|1)$ see e.g. \cite{Ivanov:2013ova,Ivanov:2013cea} and
references therein. In \cite{Ivanov:2013ova} it was shown, in
particular, that $\sSU(2|1)$ admits a supercoset which is an analog of the
harmonic analytic superspace of the standard $N = 4$, $d = 1$
supersymmetry.}
\bea
S^{3|4} =
\frac{\sSU(2|1)}{\sU(1)}~,
\label{6.34}
\eea
 which is the same as the supercoset
 $\frac{\sOSp(2|2)}{\sSO(2)}$.
A matrix realization of the generic $\sSU(2|1)$ supergroup element
that is similar to
 \eqref{6.32}
 is
\begin{equation}\label{X}
U=\left(
\begin{tabular}{c c}
 $\re^{2\ri\varphi}\sqrt{1-\theta\bar\theta}$ & $\re^{\ri\varphi}\theta^\beta$ \\
&\\
 $-\frac{\re^{2\ri\varphi} h_{\alpha}{}{^\gamma}\bar\theta_\gamma } {\sqrt{1-\theta\bar\theta}}$ & $\re^{\ri\varphi} h_{\alpha}{}^\beta $
\end{tabular}
\right )\,,
\end{equation}
where $\alpha, \b, \g=1,2$ and $\theta\bar\theta \equiv \theta^\alpha\bar\theta_\alpha=-\bar\theta\theta$ (i.e. the natural position of the index for $\theta$ is ``up" and for $\bar\theta$ is ``down").
The $2\times 2$ matrix $h_\a{}^\b$ is constrained by
\bea
 h_\alpha^{\dagger\, \gamma}h_{\gamma}{}^\beta=\delta_\alpha^\beta-\bar\theta_\alpha\theta^\beta
\qquad \rightarrow \qquad h_\alpha^{\dagger\, \beta}= (\delta_\alpha^\gamma-\bar\theta_\alpha\theta^\gamma)h_\gamma^{-1\beta}\,
\label{x}
\eea
such that
\bea
\det h =
\det h^\dagger=\frac 1{\sqrt{1-\theta\bar\theta}}\,, \quad \det (\delta_\alpha^\beta-\bar\theta_\alpha\theta^\beta)=\frac 1{1-\theta\bar\theta}=1+\theta\bar\theta+(\theta\bar\theta)^2\,.
~~~
\label{6.37}
\eea
One may check that
\bea\label{relations}
{\rm Ber} ~ U=1\, ,\qquad U^{-1}=U^\dagger~.
\eea
To check \eqref{6.37}
one should use the following identity
$
\theta^\alpha\theta_\alpha\bar\theta_\beta\bar\theta^\beta=2(\theta\bar\theta)^2
$
 and also note that
\be\label{-1}
(\delta_\alpha^\beta-\bar\theta_\alpha\theta^\beta)^{-1}=\delta_\alpha^\beta+\bar\theta_\alpha\theta^\beta(1+\theta\bar\theta)\,.
\ee
The Hermitian conjugate supermatrix is
\begin{equation}\label{Xdag}
U^\dagger=\left(
\begin{tabular}{c c}
$\re^{-2\ri\varphi}\sqrt{1-\theta\bar\theta}$ & $-\frac{\re^{-2\ri\varphi} \theta^\gamma h^\dagger_{\gamma}{}^\beta}{\sqrt{1-\theta\bar\theta}}$ \\
&\\
$\re^{-\ri\varphi}\bar\theta_\alpha$ & $\re^{-\ri\varphi} h^\dagger_{\alpha}{}^\beta $
\end{tabular}
\right )\,.
\end{equation}

It follows from \eqref{x} that
one can  define a unitary matrix $\hat h$,
$\hat h^\dagger=\hat h^{-1}$, as follows
\be\label{uni}
\hat h_\alpha{}^{\beta}=h_\alpha{}^{\gamma} (\delta_\gamma^\beta-\bar\theta_\gamma\theta^\beta)^{-\frac 12}, \qquad \sqrt{\delta_\gamma^\beta-\bar\theta_\gamma\theta^\beta}=\delta_\gamma^\beta-\frac 12\bar\theta_\gamma\theta^\beta(1+\frac 14 \theta\bar\theta)\,.
\ee

The supersphere $S^{3|4}$ is the coset \eqref{6.34}
whose element can be identified with \eqref{X} at $\varphi=0$. In this realization the Hermitian Cartan form describing the geometry of $S^{3|4}$ has the following form
\be\label{Omega}
\ri X^\dagger \rd X=\left(
\begin{tabular}{c c}
$\frac \ri 2\frac{\theta{\mathcal D}\bar\theta-{\mathcal D}\theta\,\bar\theta }{1-\theta\bar\theta}$ & $\frac{\ri{\mathcal D}\theta^\beta}{\sqrt{1-\theta\bar\theta}}$\\
&\\
$-\frac{\ri{\mathcal D}\bar\theta_\alpha}{\sqrt{1-\theta\bar\theta}}$ & $\omega_\alpha{}^\beta$
\end{tabular}
\right )\,,
\ee
where
\be\label{ED}
\omega_\alpha{}^\beta=\ri(h^\dagger \rd h)_\alpha{}^\beta
+\ri\bar\theta_\alpha \rd \theta^\beta,\qquad {\mathcal D}\theta^\alpha
=\rd\theta^\alpha +\ri\theta^\beta \omega_\beta{}^\alpha,\qquad
 {\mathcal D}\bar\theta_\alpha=\rd\bar\theta_\alpha -\ri\omega_\alpha{}^\beta\bar\theta_\beta\,.
\ee
Note that, due to the properties \eqref{x} of $h_\alpha{}^\beta$, the bosonic form $\omega_\alpha{}^\beta$ is a Hermitian matrix $\omega^\dagger=\omega$. It only depends on
$\rd\theta$ which resembles a chiral basis. The $\sSU(2|1)$ Cartan forms in a genuine chiral basis were computed in \cite{SS}.

Splitting $\omega_\alpha{}^\beta$ into traceless and traceful parts, we obtain
\be\label{ED1}
\omega_\alpha{}^\beta=E_\alpha{}^\beta+\delta_\alpha{}^\beta T,
\ee
where
\be\label{E}
E^{\alpha\beta}=\ri(h^\dagger \rd h)^{(\alpha\beta)}+\ri\bar\theta^{(\alpha} \rd\theta^{\beta)},
\ee
is the supervielbein on $S^{3|4}$ and
\be\label{U}
T=\frac 12 \omega_\alpha{}^{\alpha}
\ee
is the connection associated with the $\sU(1)$ $R$-symmetry of
$S^{3|4}$, together with the upper-left term in \eqref{Omega}, i.e.
\be\label{tildeU}
\tilde T= \frac \ri 2\frac{\theta{\mathcal D}\bar\theta-{\mathcal D}\theta\,\bar\theta }{1-\theta\bar\theta}~.
\ee
Using the unitary variables $\hat h$ defined in \eqref{uni} one can prove that $U$ and $\tilde U$ are proportional to each other and have the following form
\be\label{Uni}
\tilde T=2 T=\omega_\alpha{}^\alpha=\frac \ri 2(\theta \rd\bar\theta-\rd\theta\bar\theta)
+\ri\theta(\hat h^\dagger \rd\hat h)\bar\theta=\frac \ri 2(\theta\hat{\mathcal D}\bar\theta-\hat{\mathcal D}\theta\bar\theta),
\ee
where $\hat {\mathcal D}\bar\theta=\rd\bar\theta+(\hat h^\dagger\rd\hat h)\bar\theta$ and $\hat {\mathcal D}\theta=d\theta-\theta(\hat h^\dagger \rd\hat h)$. Note that $\tr{(\hat h^\dagger d\hat h)}=0$.

In the unitary $\hat h$-basis for the Cartan form \eqref{ED},
the expression for $\omega_\alpha{}^\beta$
becomes
\bea
\omega_\alpha{}^\beta&=&\ri(\hat h^\dagger \rd\hat h) _{\alpha}{}^{\beta}
-\frac \ri 2 (\hat{\mathcal D}\bar\theta_{\alpha}\theta^{\beta}
-\bar\theta_{\alpha}\hat{\mathcal D}\theta^{\beta})
+\frac \ri 8\left[\bar\theta_\alpha\theta^\gamma \rd(\bar\theta_\gamma\theta^\beta)
-\rd(\bar\theta_\alpha\theta^\gamma) \bar\theta_\gamma\theta^\beta\right]
\non \\
&=&\ri(\hat h^\dagger \rd\hat h) _{\alpha}{}^{\beta}-\frac \ri 2 \left(\delta_\alpha^{\alpha'}
(1+\frac 14\theta\bar\theta)-\frac 14\bar\theta_{\alpha}\theta^{\alpha'}\right)(\hat{\mathcal D}\bar\theta_{\alpha'}\theta^{\beta'} \non \\
&&-\bar\theta_{\alpha'}\hat{\mathcal D}\theta^{\beta'})\left(\delta_{\beta'}^{\beta}(1+\frac 14\theta\bar\theta)-\frac 14\bar\theta_{\beta'}\theta^{\beta}\right)~.
\label{EDhat}
\eea

The off-diagonal elements of the matrix \eqref{Omega} are the fermionic vielbeins on
$S^{3|4}$
\be\label{fE}
E^\alpha=\frac{\ri{\mathcal D}\theta^\alpha}{\sqrt{1-\theta\bar\theta}}, \qquad \bar E_\alpha=-\frac{\ri{\mathcal D}\bar\theta_\alpha}{\sqrt{1-\theta\bar\theta}}\,.
\ee


\section{Concluding comments and outlook}

In this paper we have described the supersphere $S^{3|4n}$ as
the three-dimensional  $\mathcal N=2n$ extended conformal superspace.
The  superconformal group
$\sOSp(2n|2,2)$ acts transitively on $S^{3|4n}$ by fractional linear transformations,
which at most scale the super-metric  \eqref{3.55} being invariant under the  $\sOSp(2n|2) \times \sSU(2)$ subgroup of $\sOSp(2n|2,2)$.
The supertwistor and bi-supertwistor realizations
for $S^{3|4n}$ developed in our paper provide all necessary prerequisites
for setting up a program to compute
correlations functions in off-shell superconformal field theories on $S^3$
in a way similar to the superspace approaches pursued in
\cite{Osborn,Park4,KT,Park3} or
in more recent publications \cite{Goldberger:2011yp,Maio,Goldberger:2012xb,Fitzpatrick:2014oza,Khandker:2014mpa}  which are built on
the 4D bi-supertwistor construction introduced by Siegel
\cite{Siegel93,Siegel95} and fully elaborated in \cite{K12}.\footnote{The
bi-supertwistor construction of 4D compactified Minkowski (or conformal)
superspaces was called ``superembedding formalism'' in
\cite{Goldberger:2011yp,Maio,Goldberger:2012xb}.
Indeed, this construction may be viewed as a specific example of
a general (super)embedding approach
reviewed in \cite{Sorokin} in application to  superbranes.
We also point out that there exists an  alternative use
of the name ``conformal superspace''
for the off-shell supergravity formulations developed in \cite{Butter}.
}

A natural interesting issue for further consideration is to elaborate on peculiarities and implications of the supersymmetric and superconformal structure of Wick-rotated $\mathcal N$--extended supersymmetric gauge theories such as the $\mathcal N=4$ Gaiotto-Witten models \cite{Gaiotto:2008sd} and the $\mathcal N=6$ ABJM model \cite{Aharony:2008ug} put on the $S^3$ sphere. For instance, in Minkowski space the superconformal group of the ABJM model is $\sOSp(6|4,{\mathbb R})$, while in the $3D$ space of Euclidean signature its counterpart is the supergroup $\sOSp(6|2,2)$ whose $R$-symmetry subgroup $\sSO^*(6)\simeq \sSU(3,1)$ is non-compact in contrast to the compact $R$-symmetry $\sSO(6)\simeq \sSU(4)$ of the theory in the Minkowski space. The two superconformal groups are different real forms of the complex supergroup $\sOSp(6|4,\mathbb C)$. Analogously,  the $R$-symmetry group of the Euclidean $\mathcal N=4$ Gaiotto-Witten models should be $\sSO^*(4)\simeq \sSL(2,\mathbb R)\times \sSU(2)$ for these models to be invariant under the superconformal group $\sOSp(4|2,2)$.

It is known that the harmonic \cite{GIKOS,GIOS} and projective \cite{KLR,LR}
superspace approaches are most suitable for the construction of
 supersymmetric theories with eight supercharges in four, five and six
 space-time dimensions.
 Such superspaces are obtained by extending Minkowski superspace
 by auxiliary bosonic dimensions parametrizing a coset space of the compact
 $R$-symmetry group.
 In superspaces of Euclidean signature,
  $R$-symmetry groups are often non-compact, as is the $D=3$
   $R$-symmetry group $\sSO^* (2n)$ (with
   $n >1$) considered in this paper. It is of interest to develop
 harmonic/projective superspace approaches to extended supersymmetric
 theories on $S^3$. The relevant mathematical formalism is sketched in
  appendix C. One of the most interesting cases is $\cN=4$.
  Although the corresponding $R$-symmetry group is non-compact,
  $\sSO^*(4)\simeq \sSL(2,\mathbb R)\times \sSU(2)$,  it possesses
  a compact coset space $S^1 \times S^2$ that may be used to
  define nontrivial off-shell supermultiplets.
This seems to be the right superspace setting in order
to construct Euclidean analogs of the most general  off-shell $3D$ $\cN=4$
superconformal nonlinear  $\s$-models  \cite{KPT-MvU}.

${}$
\\
\noindent
{\bf Acknowledgements:}\\
The authors are grateful to Igor Samsonov for stimulating discussions.
SMK is also grateful to Joseph Novak for reading the manuscript.
The work of SMK is supported in part by the Australian Research Council
projects DP1096372  and DP140103925. Work of DS was partially supported by the Padova University Project CPDA119349 and the INFN Special Initiative ST\&FI.
SMK  is thankful to INFN, Padova Section and
the Department of Physics and Astronomy ``Galileo Galilei'' at the
University of Padova for kind hospitality at the initial stage of this project.
D.S. would also like to acknowledge warm hospitality extended to him at the School of Physics of the University of Western Australian during a work-in-progress period.

\appendix

\section{Matrix realizations of  $ \sSp (2n, {\mathbb R} )$ and $\sSO^*(2n)$}
\label{App0}

Consider the complex symplectic group $ \sSp (2n, {\mathbb C} )$,
\bea
\sSp (2n, {\mathbb C})  := \left\{ g \in \sGL (2n,{\mathbb C}) ~,
\quad g^{\rm T} J_{n,n} g = J_{n,n}~,
\quad
J_{n,n} =\left(
\begin{array}{cc}
0   & {\mathbbm 1}_n   \\
-{\mathbbm 1}_n  &     0
\end{array}
\right) \right\}~,~~~
\label{Sp-A1}
\eea
and its subgroup $ \sSp (2n, {\mathbb R} )$ consisting of all
real symplectic matrices.\footnote{All symplectic matrices are unimodular,
$\sSp (2n, {\mathbb C}) \subset \sSL (2n, {\mathbb C})$.}
For the latter group,
there exists a different realization that is used in many applications,
see, e.g., \cite{AKL}. It is based on the isomorphism
\bea
\sSp (2n, {\mathbb R}) \cong   \sSp (2n, {\mathbb C})   \bigcap \sSU (n,n)~,
\label{Isomorphism}
\eea
where the pseudo-unitary group $\sSU(n,n) $ is defined by
\bea
\sSU(n,n) := \left\{ g \in \sSL (2n,{\mathbb C}) ~,
\quad g^\dagger I_{n,n} g = I_{n,n}~,
\quad
I_{n,n} =\left(
\begin{array}{cc}
{\mathbbm 1}_n   & 0  \\
0 &     -{\mathbbm 1}_n
\end{array}
\right) \right\}~.
\eea
To prove \eqref{Isomorphism} one performs the similarity transformation of an $\sSp (2n, {\mathbb R})$ matrix
\bea
g ~\to h := T g T^{-1} ~, \qquad g \in  \sSp (2n, {\mathbb R} )~,
\eea
where
\bea
T=  \frac{1}{\sqrt{2}}  \left(
\begin{array}{cc}
{\mathbbm 1}_n   ~& \ri {\mathbbm 1}_n   \\
\ri {\mathbbm 1}_n  ~&     {\mathbbm 1}_n
\end{array}
\right) ~.
\label{T-matrix}
\eea
This matrix is symmetric and unitary, $T^\dagger T={\mathbbm 1}_{2n} $, and such that
$TJ_{n,n}T = J_{n,n}$ and $TJ_{n,n} T^{-1} = -\ri I_{n,n}$.

Consider now the group
\bea
\sSO^*(2n) = \sSO(2n, {\mathbb C})    \bigcap\sSp (2n, {\mathbb C})
:= \Big\{ g \in \sSp (2n, {\mathbb C}) ~, \quad g^{\rm T} g = {\mathbbm 1}_{2n}
\Big\}
~,
\label{AA.6}
\eea
with $\sSp (2n, {\mathbb C}) $ defined by \eqref{Sp-A1}.
This group is isomorphic to
\bea
H := \left\{ h \in \sSU (n,n) ~, \quad h^{\rm T} I_{n,n} J_{n,n} h = I_{n,n}J_{n,n} ~,
\quad I_{n,n} J_{n,n}
=\left(
\begin{array}{cc}
0   & {\mathbbm 1}_n   \\
{\mathbbm 1}_n  &     0
\end{array}
\right)
 \right\} ~.~~~
\eea
The proof is based on considering the similarity transformation
\bea
g ~\to h := T g T^{-1} ~, \qquad g \in  \sSO^*(2n)~,
\eea
with the matrix $T$ given by \eqref{T-matrix}.

\section{Conformal spaces}\label{AppendixA}

Consider a $d$-dimensional pseudo-Euclidean space ${\mathbb E}^{s, t}$, 
$d=s+t$,
parametrized by Cartesian coordinates
$x^{a}$,    where $a = 1,\dots,  d$,
and endowed with the metric
\bea
\eta_{a b} = {\rm diag}  (1, \dots, 1,  -1,\dots, -1)~,
\eea
with $s>0$ `pluses' and $t$ `minuses' on the diagonal.
The conformal algebra of  ${\mathbb E}^{s, t}$ is known to be ${\frak so} (1+s , 1+t)$.
It is also known that the corresponding conformal group
does not act globally on ${\mathbb E}^{s, t}$.
Its action is well defined on a conformal compactification $\overline{\mathbb E}^{s,t}$
of ${\mathbb E}^{s, t}$.
Similar to the works of Veblen \cite{Veblen} and Dirac \cite{Dirac},
the space $\overline{\mathbb E}^{s,t}$ may be introduced as follows.
We consider  a $(d+2)$-dimensional
pseudo-Euclidean space ${\mathbb E}^{1+s, 1+1}$
with coordinates $X^\ha = (X^{-1}, X^a, X^{d+1})$
and metric
\bea
\eta_{\ha \hb}  =\left(
\begin{array}{ c c c}
1  & 0 &  0\\
0 & \eta_{ab} &0 \\
 0 & 0 & -1\\
\end{array}
\right)~.
\eea
Embedded into ${\mathbb E}^{1+s,1+t} $ is the cone $\cC$
defined by
\bea
\eta_{\ha \hb} X^{\ha} X^{\hb} = 0~.
\label{cone}
\eea
By definition, $\overline{\mathbb E}^{s,t}$
is the space of all straight lines belonging to  $\cC$ and passing through the origin of
 $ {\mathbb E}^{1+s,1+t}$.
 It can be defined as the quotient space of $\cC \setminus \{ 0 \} $ with respect to  the equivalence relation
\bea
X^{\ha} ~\sim ~ \l \, X^{\ha} ~, \qquad \l \in {\mathbb R} \setminus \{ 0 \}~,
\label{eqrelation}
\eea
which identifies all points on a straight line in $ {\mathbb E}^{1+s, 1+t}$.
The group $\sO(1+s, 1+t)$ naturally acts on  $\overline{\mathbb E}^{s,t}$
such that  the group elements $g$ and $-g$
generate the same transformation,  for any $g \in  \sO(1+s, 1+t)$.
The conformal group of $ {\mathbb E}^{s, t}$, $\mathsf{Conf} ({\mathbb E}^{s,t})$,
 is defined to be
   $\sO(1+s , 1+t)/{\mathbb Z}_2$.
If $d$ is odd, 
the conformal group may be identified with $\sSO(1+s , 1+t)$.
The  space
$\overline{\mathbb E}^{s,t}$
is a homogeneous space of
$\mathsf{Conf} ({\mathbb E}^{s,t})$.

As a topological space,
$\overline{\mathbb E}^{s,t}$
is homeomorphic to
\begin{subequations}
\bea
\overline{\mathbb E}^{s,t}
&=& (S^{s} \times S^t)/{\mathbb Z}_2~, \qquad t>0~;
\label{A.5a}\\
\overline{\mathbb E}^{d}
&  \equiv  &
\overline{\mathbb E}^{d,0}
= S^{d} ~.
\eea
\end{subequations}
Indeed, for $t>0$
the constraint \eqref{cone} and equivalence relation
\eqref{eqrelation} can be used to choose $X^{\au}$ such that
\bea
(X^{-1})^2 + \sum_{i=1}^s(X^{i})^2 = \sum_{i=s+1}^{d} (X^i)^2  + (X^{d+1})^2= 1~.
\eea
For such a choice, the equivalence relation \eqref{eqrelation}
still allows us to identify
$X^{\ha}$ and $-X^{\ha}$, which is the reason for  ${\mathbb Z}_2$  in
\eqref{A.5a}. When $t=0$, we have $X^{d+1} \neq 0$ for any non-zero point on
the cone $\cC$. As a result,  the equivalence relation \eqref{eqrelation}
can be used to choose $X^{d+1}=1$, which means
\bea
(X^{-1})^2 + \sum_{i=1}^s(X^{i})^2 =1~.
\eea

Pseudo-Euclidean space $ {\mathbb E}^{s, t}$
 can be identified, e.g.,
with the open {\it dense} domain $U_+$ of
$\overline{\mathbb E}^{s,t}$
on which $X^{-1}+ X^{d+1}  \neq 0$.
This domain can be parametrized by  {\it inhomogeneous} coordinates
\bea
x^a = \frac{X^a}{X^{-1}  + X^{d+1}}~,
\eea
which are invariant under the identification \eqref{eqrelation}.
In terms of these coordinates, one obtains a standard action of the conformal group
in  ${\mathbb E}^{s,t}$. Along with $U_+$, we can consider the open set
$U_-$ of
$\overline{\mathbb E}^{s,t}$
 on which $X^{-1}-X^{d+1}   \neq 0$.
The latter may be parametrized by coordinates
\bea
y^a = \frac{X^a}{X^{-1}  - X^{d+1}}~.
\eea
In the overlap of the two charts, $U_+ \bigcap U_-$, it holds that
\bea
y^a = - \frac{x^a}{x^2} ~, \qquad x^2 = \eta_{ab}x^a x^b ~.
\eea
In the Euclidean case, $t=0$, the charts $U_+ $ and $ U_-$
constitute an atlas of the conformal space, $S^d = U_+ \bigcup U_-$.

If at least one of the dimensions $s$ and $t$ is even,
the conformal group consists of two disjoint connected components,
\bea
\mathsf{Conf} ({\mathbb E}^{s,t}) \cong \sSO_0 (1+s , 1+t) \bigcup
I \cdot \sSO_0(1+s , 1+t)~,
\eea
where $I$ is a discrete transformation. If $t=0$,  $I$ may be defined
as follows $I: X^{-1} \to -X^{-1}$, $X^a \to X^a$, $X^{d+1} \to X^{d+1}$.
This {\it conformal inversion} acts on
$\overline{\mathbb E}^{s,t}$
as
\bea
x^a ~\to ~ \frac{x^a}{x^2}~,
\eea
and it does not 
belong to the connected component of  the identity of the conformal group
in the Euclidean case, $t=0$.


\section{Fibre bundles over the supersphere}

It is possible to introduce fibre bundles over $S^{3|4n}$
by generalizing the construction of subsection \ref{Subsection3.2}
to include odd supertwistors.\footnote{Our approach in this appendix is inspired
 by the construction of compactified harmonic/projective superspaces
 with Lorentzian signature given in \cite{KPT-MvU,K06,K12}. These papers built on
earlier works \cite{Rosly,LN2,HH}. }
Odd supertwistors will parametrize fibres
over the supersphere. Given such an odd supertwistor $\J$, it is defined
by the following two conditions: (i) it is orthogonal
to the even supertwistors $T^\m$ parametrizing $S^{3|4n}$
with respect to the inner products \eqref{3.7},
\bea
\langle T^\m | \J \rangle_\X = 0~,\qquad
\langle T^\m | \J \rangle_\U = 0~;
\label{C.1}
\eea
(ii) it is defined modulo the equivalence relation
\bea
\J ~\sim ~ \J + T^\m a_\m~,
\label{C.2}
\eea
for arbitrary $a$-numbers $a_\m$ (i.e. odd elements of the Grassmann algebra).
When $T^\m$ are chosen as in \eqref{3.18}, the equivalence relation \eqref{C.2}
allows us to choose $\J$ to be
\bea
\J = \left(
\begin{array}{c}
 v_i  \\  \hline
  \x_\a  \\
 0
\end{array}
\right)~,
\label{C.3}
\eea
where $v_i$ is an even $2n$-vector, and  $\x_\a$ an odd two-spinor.
Imposing the orthogonality conditions \eqref{C.1} gives, respectively,
\bea
\x &=& -(\bm h^\dagger)^{-1} \Q^\dagger\O v \\
&=& \s_2 (\bm h^{\rm T} )^{-1} \Q^{\rm T} v~.
\eea
These two expressions for $\x$ are actually equivalent due to the reality conditions
\eqref{3.20}. We see that $\J$ brings in only bosonic degrees of freedom that are described
by the complex $2n$-vector  $v_i$.
By taking several odd supertwistors and imposing $\sOSp (2n|2,2)$ invariant conditions,
the bosonic $v$-variables may be made to parametrize a homogeneous space
of $\sSO^*(2n)$.

In the case of a single odd supertwistor, we may impose the following conditions
\bea
\langle \J | \J \rangle_\X = 0~,\qquad
\langle \J | \J \rangle_\U = 0~.
\eea
It is easy to see that for $n=1$ the $v$-variables describe a one-sphere $S^1$.

Given several odd supertwistors $\J^M$, with $M=1,\dots, m$, we may choose them
to describe odd $m$ planes. Then the equivalence relation \eqref{C.2} should be
replaced by a more general one of
the form
\bea
\J^M ~\sim ~ \J^N{}A_N{}^M + T^\m a_\m{}^M~, \qquad
A=(A_M{}^N)\in \sGL (m, {\mathbb C})~.
\label{C.7}
\eea
Now we may impose $\sOSp (2n|2,2)$ invariant conditions
in terms of the supermatrix $\hat \J := (\J_A{}^M)$.
In particular, for $n>1$ and $m=2$ we may choose the conditions
 \bea
 \hat{ \J}{}^\dagger \X \hat \J > 0~, \qquad \hat{\J}{}^{\rm T} \U \hat \J =0~,
 \eea
 where the notation $ \hat{ \J}{}^\dagger \X \hat \J > 0$ means
 that the Hermitian matrix $ \hat{ \J}{}^\dagger \X \hat \J $ is positive definite.
 For this choice the $v$-variables describe the Hermitian symmetric space
 $\sSO^*(2n) /\sU(n)$, see, e.g., \cite{AKL}.
In the extreme case $m =2n$, no degrees of freedom
are described by the $v$-variables,
since the equivalence relation \eqref{C.7} allows us to bring any odd $2n$-plane to the form
 \bea
 \hat \J =
  \left(
\begin{array}{c}
 \O
 \\  \hline
 -(\bm h^\dagger)^{-1} \Q^\dagger
 \\
 0
\end{array}
\right)~.
\eea
One may check that this odd $2n$-plane is real under the star-map
\eqref{3.13}.

\begin{footnotesize}

\end{footnotesize}


\begin{thebibliography}{66}

\bibitem{KWY}
  A.~Kapustin, B.~Willett and I.~Yaakov,
  ``Exact results for Wilson loops in superconformal Chern-Simons theories with matter,''
  JHEP {\bf 1003}, 089 (2010)
  [arXiv:0909.4559 [hep-th]].

\bibitem{Jafferis}
  D.~L.~Jafferis,
  ``The exact superconformal R-symmetry extremizes Z,''
  JHEP {\bf 1205}, 159 (2012)
  [arXiv:1012.3210 [hep-th]].

\bibitem{Hama:2010av}
  N.~Hama, K.~Hosomichi and S.~Lee,
  ``Notes on SUSY gauge theories on three-sphere,''
  JHEP {\bf 1103}, 127 (2011)
  [arXiv:1012.3512 [hep-th]].

\bibitem{SS}
I.~B.~Samsonov and D.~Sorokin,
``Superfield theories on $S^3$ and their localization,''
JHEP {\bf 1404}, 102 (2014)
[arXiv:1401.7952 [hep-th]].

\bibitem{KT-M11}
S.~M.~Kuzenko and G.~Tartaglino-Mazzucchelli,
 ``Three-dimensional N=2 (AdS) supergravity and associated supercurrents,''
JHEP {\bf 1112}, 052 (2011)
[arXiv:1109.0496 [hep-th]].


\bibitem{KLT-M12}
S.~M.~Kuzenko, U.~Lindstr\"om and G.~Tartaglino-Mazzucchelli,
``Three-dimensional (p,q) AdS superspaces and matter couplings,''
JHEP {\bf 1208}, 024 (2012)
[arXiv:1205.4622 [hep-th]].

\bibitem{BKT-M}
D.~Butter, S.~M.~Kuzenko and G.~Tartaglino-Mazzucchelli,
``Nonlinear sigma models with AdS supersymmetry in three dimensions,''
JHEP {\bf 1302}, 121 (2013)
[arXiv:1210.5906 [hep-th]].

\bibitem{KT-M14}
  S.~M.~Kuzenko and G.~Tartaglino-Mazzucchelli,
  ``N = 4 supersymmetric Yang-Mills theories in $AdS_3$,''
  JHEP {\bf 1405}, 018 (2014)
  [arXiv:1402.3961 [hep-th]].


\bibitem{KPT-MvU}
S.~M.~Kuzenko, J.~-H.~Park, G.~Tartaglino-Mazzucchelli and R.~Unge,
``Off-shell superconformal nonlinear sigma-models in three dimensions,''
JHEP {\bf 1101}, 146 (2011)
  [arXiv:1011.5727 [hep-th]].

\bibitem{K06}
 S.~M.~Kuzenko,
``On compactified harmonic/projective superspace, 5D superconformal
theories, and all that,''
Nucl.\ Phys.\  B {\bf 745}, 176 (2006)
[arXiv:hep-th/0601177].

\bibitem{K12}
  S.~M.~Kuzenko,
  ``Conformally compactified Minkowski superspaces revisited,''
  JHEP {\bf 1210}, 135 (2012)
  [arXiv:1206.3940 [hep-th]].

\bibitem{Lukierski:1981ht}
  J.~Lukierski and A.~Nowicki,
  ``Superspinors and graded Lorentz groups in three dimensions, four dimensions and five dimensions,''
  Fortsch.\ Phys.\  {\bf 30}, 75 (1982).

\bibitem{Lukierski:1983qc}
  J.~Lukierski and A.~Nowicki,
  ``Quaternionic supergroups and $D=4$ Euclidean extended supersymmetries,''
  Annals Phys.\  {\bf 166}, 164 (1986).

\bibitem{PvNS}
  K.~Pilch, P.~van Nieuwenhuizen and M.~F.~Sohnius,
  ``De Sitter superalgebras and supergravity,''
  Commun.\ Math.\ Phys.\  {\bf 98}, 105 (1985).

\bibitem{LN}
  J.~Lukierski and A.~Nowicki,
  ``All possible de Sitter superalgebras and the presence of ghosts,''
  Phys.\ Lett.\ B {\bf 151}, 382 (1985).

  	
\bibitem{Veblen} O. Veblen, ``Geometry of four-component spinors,''
Proc. Nat. Acad. Sci. {\bf 19}, 503 (1933).

\bibitem{Dirac} P. A. M. Dirac, ``Wave equations in conformal space,''
Ann. Math. {\bf 37}, 429 (1936).



\bibitem{Ferber}
  A.~Ferber, ``Supertwistors and conformal supersymmetry,''
  Nucl.\ Phys.\ B {\bf 132}, 55 (1978).

\bibitem{Manin} Yu. I. Manin,
``Holomorphic supergeometry and Yang-Mills superfields,''
J. Sov. Math. {\bf 30}, 1927 (1985);
{\it Gauge Field Theory and Complex Geometry},
Springer, Berlin, 1988.

\bibitem{Niederle}
  M.~Kotrla and J.~Niederle,
  ``Supertwistors and superspace,''
  Czech.\ J.\ Phys.\ B {\bf 35}, 602 (1985).

\bibitem{Siegel93}
  W.~Siegel,
  ``Green-Schwarz formulation of self-dual superstring,''
  Phys.\ Rev.\ D {\bf 47}, 2512 (1993)
  [hep-th/9210008].

\bibitem{Siegel95}
  W.~Siegel,
  ``Super multi-instantons in conformal chiral superspace,''
  Phys.\ Rev.\ D {\bf 52}, 1042 (1995)
  [hep-th/9412011].



\bibitem{DeWitt}
B.~S.~DeWitt,  {\it Supermanifolds}, Cambridge University Press,
Cambridge, 1992.

\bibitem{BK}
I.~L. Buchbinder and S.~M. Kuzenko, {\it Ideas and Methods of Supersymmetry and
Supergravity, Or a Walk Through Superspace}, IOP, Bristol, 1998.

\bibitem{Cartan} E. Cartan, {\it The Theory of Spinors}, Dover Publications,
New York, 1981.

\bibitem{VA}
D.~V.~Volkov and V.~P.~Akulov,
``Possible universal neutrino interaction,''
  JETP Lett.\  {\bf 16}, 438 (1972)  [Pisma Zh.\ Eksp.\ Teor.\ Fiz.\  {\bf 16}, 621 (1972)];
  ``Is the neutrino a Goldstone particle?,''
  Phys.\ Lett.\  B {\bf 46}, 109 (1973).

\bibitem{AV}
V.~P.~Akulov and D.~V.~Volkov, ``Goldstone fields with spin 1/2,''
Theor.\ Math.\ Phys.\  {\bf 18}, 28 (1974)  [Teor.\ Mat.\ Fiz.\  {\bf 18}, 39 (1974)].



\bibitem{Knapp} A. W. Knapp, {\it Representation Theory of
Semisimple Groups}, Princeton University Press, Princeton, 2001.




\bibitem{Plyushchay:2003gv}
  M.~Plyushchay, D.~Sorokin and M.~Tsulaia,
  ``Higher spins from tensorial charges and $OSp(N| 2n)$ symmetry,''
  JHEP {\bf 0304}, 013 (2003)
  [hep-th/0301067].
	
	
\bibitem{AKL}
  M.~Arai, S.~M.~Kuzenko and U.~Lindstr\"om,
  ``Hyperk\"ahler sigma models on cotangent bundles of Hermitian symmetric
  spaces using projective superspace,''
  JHEP {\bf 0702}, 100 (2007)  [arXiv:hep-th/0612174].


\bibitem{GGRS}
 S.~J.~Gates, Jr., M.~T.~Grisaru, M.~Ro\v{c}ek and W.~Siegel,
{\it Superspace, or One Thousand and One Lessons in Supersymmetry},
Benjamin/Cummings, Reading, MA,  1983, hep-th/0108200.


\bibitem{Ivanov:2013ova}
  E.~Ivanov and S.~Sidorov,
  ``Deformed supersymmetric mechanics,''
  Class.\ Quant.\ Grav.\  {\bf 31} (2014) 075013
  [arXiv:1307.7690 [hep-th]].

\bibitem{Ivanov:2013cea}
  E.~Ivanov and S.~Sidorov,
  ``Super K\"ahler oscillator from $SU(2|1)$ superspace,''
  arXiv:1312.6821 [hep-th].

\bibitem{Osborn}
  H.~Osborn,   ``N = 1 superconformal symmetry in four-dimensional
quantum field theory,''
Annals Phys.\  {\bf 272}, 243 (1999) [hep-th/9808041].

\bibitem{Park4}
 J.-H.~Park,  ``Superconformal symmetry and correlation functions,''
 Nucl.\ Phys.\ B {\bf 559}, 455 (1999)   [hep-th/9903230].

\bibitem{KT}
S.~M.~Kuzenko and S.~Theisen,
 ``Correlation functions of conserved currents in N = 2 superconformal
theory,''  Class.\ Quant.\ Grav.\  {\bf 17}, 665 (2000)  [hep-th/9907107].

\bibitem{Park3}
  J.-H.~Park,
  ``Superconformal symmetry in three-dimensions,''
  J.\ Math.\ Phys.\  {\bf 41}, 7129 (2000)
  [arXiv:hep-th/9910199].

\bibitem{Goldberger:2011yp}
  W.~D.~Goldberger, W.~Skiba and M.~Son,
  ``Superembedding methods for 4d N=1 SCFTs,''
  Phys.\ Rev.\ D {\bf 86}, 025019 (2012)
  [arXiv:1112.0325 [hep-th]].


\bibitem{Maio}
  M.~Maio,
  ``Superembedding methods for 4d N-extended SCFTs,''
  Nucl.\ Phys.\ B {\bf 864}, 141 (2012)
  [arXiv:1205.0389 [hep-th]].

\bibitem{Goldberger:2012xb}
  W.~D.~Goldberger, Z.~U.~Khandker, D.~Li and W.~Skiba,
  ``Superembedding methods for current superfields,''
  Phys.\ Rev.\ D {\bf 88}, 125010 (2013)
  [arXiv:1211.3713 [hep-th]].

\bibitem{Fitzpatrick:2014oza}
  A.~L.~Fitzpatrick, J.~Kaplan, Z.~U.~Khandker, D.~Li, D.~Poland and D.~Simmons-Duffin,
  ``Covariant approaches to superconformal blocks,''
  arXiv:1402.1167 [hep-th].

\bibitem{Khandker:2014mpa}
  Z.~U.~Khandker, D.~Li, D.~Poland and D.~Simmons-Duffin,
  ``$\mathcal{N}=1$ superconformal blocks for general scalar operators,''
  arXiv:1404.5300 [hep-th].


\bibitem{Sorokin}
  D.~P.~Sorokin,
  ``Superbranes and superembeddings,''
  Phys.\ Rept.\  {\bf 329}, 1 (2000)
  [hep-th/9906142].

\bibitem{Butter}
  D.~Butter,
  ``$\cN=1$ conformal superspace in four dimensions,''
  Annals Phys.\  {\bf 325}, 1026 (2010)
  [arXiv:0906.4399 [hep-th]];
  ``$\cN=2$ conformal superspace in four dimensions,''
  JHEP {\bf 1110}, 030 (2011)
  [arXiv:1103.5914 [hep-th]].


\bibitem{Gaiotto:2008sd}
  D.~Gaiotto and E.~Witten,
  ``Janus configurations, Chern-Simons couplings, and the theta-angle in N=4 super Yang-Mills theory,''
  JHEP {\bf 1006}, 097 (2010)
  [arXiv:0804.2907 [hep-th]].

\bibitem{Aharony:2008ug}
  O.~Aharony, O.~Bergman, D.~L.~Jafferis and J.~Maldacena,
  ``N=6 superconformal Chern-Simons-matter theories, M2-branes and their gravity duals,''
  JHEP {\bf 0810}, 091 (2008)
  [arXiv:0806.1218 [hep-th]].



\bibitem{GIKOS}
  A.~Galperin, E.~Ivanov, S.~Kalitsyn, V.~Ogievetsky and E.~Sokatchev,
  ``Unconstrained N = 2 matter, Yang-Mills and supergravity theories in harmonic
  superspace,''
  Class.\ Quant.\ Grav.\  {\bf 1}, 469 (1984).

\bibitem{GIOS}
A.~S.~Galperin, E.~A.~Ivanov, V.~I.~Ogievetsky and E.~S.~Sokatchev,
{\it Harmonic Superspace}, Cambridge University Press,  2001.

\bibitem{KLR}
A. Karlhede, U. Lindstr\"om and M. Ro\v cek,
``Self-interacting tensor multiplets in N = 2 superspace,''
Phys.\ Lett.\ B {\bf 147}, 297 (1984).

\bibitem{LR}
U.~Lindstr\"om and M.~Ro\v{c}ek,
``New hyperk\"ahler  metrics  and new supermultiplets,''
  Commun.\ Math.\ Phys.\  {\bf 115}, 21 (1988);
%
 ``N = 2 super Yang-Mills theory in projective superspace,''
Commun.\ Math.\ Phys.\  {\bf 128}, 191 (1990).


\bibitem{Rosly}
  A.~A.~Rosly,
 ``Gauge fields in superspace and twistors,''
  Class.\ Quant.\ Grav.\  {\bf 2} (1985) 693.

\bibitem{LN2}
  J.~Lukierski and A.~Nowicki,
  ``General superspaces from supertwistors,''
  Phys.\ Lett.\ B {\bf 211} (1988) 276.

\bibitem{HH}
  P.~S.~Howe and G.~G.~Hartwell,
  ``A superspace survey,''
  Class.\ Quant.\ Grav.\  {\bf 12} (1995) 1823.


\end{thebibliography}
\end{document}